\def\Z{{\bf Z}}
\def\Wedge{{\wedge}}
\documentstyle[12pt]{article}
\parskip=\medskipamount                 
\thicklines                     
\newcommand{\bimn}[7]{\bibitem{#1}#2,
{\em #3},
{ #4}\hspace{0.25em}{\bf
#5}\hspace{0.25em}(#6)\hspace{0.25em}{#7}.}

\def\inbar{\vrule height1.5ex width.4pt depth0pt}
\def\IC{\relax\,\hbox{$\inbar\kern-.3em{\rm C}$}}
\def\IN{\relax{\rm I\kern-.18em N}}
\def\IQ{\relax\,\hbox{$\inbar\kern-.3em{\rm Q}$}}
\def\IR{\relax{\rm I\kern-.18em R}}
\def\ZZ{\relax{\sf Z\kern-.4em Z}}

   \def\cD{{\cal D}}
  \def\cH{{\cal H}} 
 \def\cK{{\cal K}} \def\cL{{\cal L}} 
 \def\cO{{\cal O}}  
 
\catcode`\@=11

\newif\if@fewtab\@fewtabtrue

\catcode`\@=11

\newif\if@fewtab\@fewtabtrue

{\count255=\time\divide\count255 by 60
\xdef\hourmin{\number\count255}
\multiply\count255 by-60\advance\count255 by\time
\xdef\hourmin{\hourmin:\ifnum\count255<10 0\fi\the\count255}}
\def\ps@draft{\let\@mkboth\@gobbletwo
    \def\@oddhead{}
    \def\@oddfoot
      {\hbox to 7 cm{\footnotesize {\em Draft of \jobname:} \draftdate
       \hfil}\hskip -7cm\hfil\rm\thepage \hfil}
    \def\@evenhead{}\let\@evenfoot\@oddfoot}

\def\ceqno{\global\@fewtabfalse
    \ifcase\@eqcnt \def\@tempa{& & &}\or \def\@tempa{& &}
      \or \def\@tempa{&}
      \or\def\@tempa{}\fi\@tempa
{\rm(\theequation)}}

\def\aeqno#1{\global\@fewtabfalse
    \ifcase\@eqcnt \def\@tempa{& & &}\or \def\@tempa{& &}
      \or \def\@tempa{&}
      \or\def\@tempa{}\fi\@tempa
{\rm(\theequation,#1)}}

\def\label#1{\ifnum\draftcontrol=1
 \global\def\draftnote{$\scriptstyle #1$}\fi
 \@bsphack\if@filesw {\let\thepage\relax
   \def\protect{\noexpand\noexpand\noexpand}%
\xdef\@gtempa{\write\@auxout{\string
      \newlabel{#1}{{\@currentlabel}{\thepage}}}}}\@gtempa
   \if@nobreak \ifvmode\nobreak\fi\fi\fi
  \@esphack}
\def\C{{\bf C}}
\def\alabel#1#2{\label{#1}\global\@fewtabfalse
    \ifcase\@eqcnt \def\@tempa{& & &}\or \def\@tempa{& &}
      \or \def\@tempa{&}
      \or\def\@tempa{}\fi\@tempa
{\hbox to 3cm{\phantom{\rm(\theequation,#2)}
\draftnote \hfil}\hskip -3cm {\rm(\theequation,#2)}}}

\def\clabel#1{\label{#1}\global\@fewtabfalse
    \ifcase\@eqcnt \def\@tempa{& & &}\or \def\@tempa{& &}
      \or \def\@tempa{&}
      \or\def\@tempa{}\fi\@tempa
{\hbox to 3cm{\phantom{\rm(\theequation)}
\draftnote \hfil}\hskip -3cm{\rm(\theequation)}}}

\def\eqnarray{\def\draftnote{{}}\global\@fewtabtrue
\stepcounter{equation}\let\@currentlabel=\theequation
\global\@eqnswtrue
\global\@eqcnt\z@\tabskip\@centering\let\\=\@eqncr
$$\halign to \displaywidth\bgroup\@eqnsel\hskip\@centering\@eqcnt\z@
  $\displaystyle\tabskip\z@{##}$&\global\@eqcnt\@ne
  \hskip 1\arraycolsep \hfil$\displaystyle{##}$\hfil
  &\global\@eqcnt\tw@ \hskip 1\arraycolsep
$\displaystyle\tabskip\z@{##}$
\hfil  \tabskip\@centering&\global\@eqcnt\thr@@\llap{##}\tabskip\z@
\cr}

\def\endeqnarray{\@@eqncr\egroup
      \global\advance\c@equation\m@ne$$\global\@ignoretrue}

\def\@eqnnum{\hbox to 3cm{\phantom{\rm(\theequation)} \draftnote
                         \hfil}\hskip -3cm {\rm(\theequation)}}

\def\@@eqncr{\let\@tempa\relax
    \ifcase\@eqcnt \def\@tempa{& & &}\or \def\@tempa{& &}
      \or \def\@tempa{&}
      \or\def\@tempa{}
\fi\@tempa
\if@eqnsw
\if@fewtab\@eqnnum\fi
\stepcounter{equation}\fi\global
\@eqnswtrue\global\@eqcnt\z@\global\@fewtabtrue\cr}

\def\draftcite#1{\ifnum\draftcontrol=1#1\else{}\fi}

\def\@lbibitem[#1]#2{\item{}\hskip -3cm \hbox to 2cm
{\hfil$\scriptstyle\draftcite{#2}$}\hskip
1cm[\@biblabel{#1}]\if@filesw
     {\def\protect##1{\string ##1\space}\immediate
      \write\@auxout{\string\bibcite{#2}{#1}}}\fi\ignorespaces}

\def\@bibitem#1{\item\hskip -3cm \hbox to 2cm
{\hfil $\scriptstyle\draftcite{#1}$}\hskip 1cm
\if@filesw \immediate\write\@auxout
       {\string\bibcite{#1}{\the\value{\@listctr}}}\fi\ignorespaces}

\def\nsection#1{\section{#1}\setcounter{equation}{0}}
\def\nappendix#1{\def\thesection{A#1}\section*{Appendix #1}
\def\theequation{{A#1.\arabic{equation}}}
\def\theproposition{{A#1.\arabic{proposition}}}
\setcounter{equation}{0}
\setcounter{proposition}{0}}

\def\draftdate{\number\month/\number\day/\number\year\ \ \ \hourmin }

\global\def\draftcontrol{0}
\catcode`\@=12

\def\theequation{{\thesection.\arabic{equation}}}

\def\qq{\begin{eqnarray}}
\def\qqq{\end{eqnarray}}
\def\rx#1{~(\ref{#1})}
\def\ex#1{eq.\rx{#1}}
\def\eex#1{eqs.\rx{#1}}
\def\cx#1{~\cite{#1}}
\def\rw#1{~\ref{#1}}
\def\nn{\begin{eqnarray}}
\def\nnn{\end{eqnarray}}

\def\ie{{ i.e.\ }}
\def\eg{{ e.g.\ }}
\def\cf{{ cf.\ }}
\def\rhs{right hand side }
\def\lhs{left hand side }

\def\Tr{\mathop{{\rm Tr}}\nolimits}
\def\Vol{\mathop{{\rm Vol}}\nolimits}

\def\ordH{{|H_1(M,\ZZ)|} }
\def\ordHa#1{ { | H_1(#1,\ZZ) |} }

\def\p{^{\prime}}
\def\pp{^{\prime\prime}}

\def\prosign{\mathop{{\rm sign}}\nolimits}

\def\Pexp{\mathop{{\rm Pexp}}\nolimits}

\def\lrbc#1{ \left( #1 \right) }
\def\lrbs#1{ \left[ #1 \right] }

\def\lrl#1{ \langle #1 \rangle }

\def\sc#1#2{ \langle #1 | #2 \rangle }
\def\sco#1#2#3{ \langle #1 | #2 | #3 \rangle }
\def\bar{\overline}

\def\hk{ hyper-K\"{a}hler }
\def\k{ K\"{a}hler }

\def\phib{ \bar{\phi} }
\def\vphi{ \varphi }
\def\vphib{ \bar{\vphi} }
\def\cvphi{ \vphi_0 }

\def\eps{ \epsilon }
\def\epsn{ \epsilon }
\def\veps{ \varepsilon }

\def\zphim{ Z( M;\phi_0^i) }
\def\zz{ Z_0 (M;\cvphi^i ) }

\def\zf{ Z_{\eta\chi\varphi} (M,\hx;\cvphi^i ) }

\def\zgmh{ Z_\Gd(M,\hx;\cvphi^i) }
\def\zamlh{ Z_{\wa_1,\ldots,\wa_N}(M,L,\hx) }

\def\igi{ I_{\Gd, a} }

\def\wgh{ W_\Gd(\hx;\cvphi^i) }

\def\pp#1#2{ { \partial{#1} \over \partial{#2} } }

\def\bc{ \bar{c} }
\def\bI{ \bar{I} }
\def\bJ{ \bar{J} }
\def\bK{ \bar{K} }
\def\bL{ \bar{L} }
\def\bM{ \bar{M} }
\def\bN{ \bar{N} }
\def\bQ{ \bar{Q} }

\def\iphi{ \cD \phi^i }
\def\ieta{ \cD \eta^I }
\def\ichi{ \cD \chi^I_\mu }
\def\pint{ \int \iphi \ieta \ichi }
\def\iA{ \cD A^a_\mu }

\def\Gphi{ G^{(\vphi)} (x_1,x_2) }
\def\Gchi#1{ G^{(\chi)}_{#1} (x_1,x_2) }
\def\Gt#1{ \tilde{G}_{#1} (x_1,x_2) }

\def\Scs{ S_{\rm CS} }
\def\Scsc{ S_{\rm CS}^{(c)} }
\def\Scsgf{ S_{\rm CS,gf} }
\def\zcs{ Z_{\rm CS} (M;k) }
\def\zcsc{ Z_{\rm CS}^{(c)} (M;k) }
\def\ztrcs{ Z_{\rm CS}^{\rm (tr)} (M;k)}
\def\ztrocs{ Z_{\rm CS}^{\rm (tr,0)} (M;k)}

\def\he{  {1\over \sqrt{h} } \epsilon^{\mu\nu\rho} }
\def\hdx{ \sqrt{h} \, d^3x }
\def\gdphi{ \sqrt{g} \, d^{4n} \cvphi^i }

\def\tG{ \tilde{G} }
\def\hbcs{ \hbar_{\rm CS} }

\def\intr{ \int_{[{\rm triv}]} }

\def\dh{ b_1 }

\def\lmax{ \Lambda^{\rm max} }

\def\vocs{ V_{1,{\rm CS} } }
\def\vtcs{ V_{2,{\rm CS} } }

\def\ah{ {X_{\rm AH} } }
\def\hx{ X }

\def\fr{ {\rm fr} }
\def\ihfr#1{ I(h,#1) }
\def\dfr{ \Delta\fr }

\def\lcw{ \lambda_{\rm CW} }
\def\lc{ \lambda_{\rm C} }

\def\sdim{\mathop{{\rm sdim}}\nolimits}

\def\jacbi{ \Omega_{IKLS}\Omega_{JMNT}+\Omega_{IKMS}\Omega_{JNLT}
     + \Omega_{IKNS}\Omega_{JLMT} }

\def\ZZ{ {\bf Z} }
\def\IR{ {\bf R} }

\def\Gs{ \Gamma }
\def\Gd{ \Gamma }

\def\bg{ b_\Gd }
\def\smg{ \sigma }
\def\sg{ s }
\def\ia{ \alpha }
\def\ib{ \beta }
\def\ipe{ l }
\def\ipc{ l }
\def\iqc{ m }

\def\iq{ m }
\def\t{ t }

\def\sA{ {\cal A} }

\def\wa{ \alpha }
\def\Va{ {\cal V}_\wa }
\def\Vab{ {\cal V}_{\wa_b} }

\def\intcycl{ \int_{\rm cycl.\,ord.} }

\def\omx{ \omega }
\def\thetx{ \theta }
\def\cb#1{ \chif{#1} }
\def\tx{ u }
\def\treid{ \tau_{\rm R} (M;e^{2i\tx}) }
\def\treido{ \tau_{\rm R} (M;e^{i\tx}) }
\def\treidmk{ \tau_{\rm R} (\mmk;e^{2i\tx}) }
\def\treidomk{ \tau_{\rm R} (\mmk;e^{i\tx}) }
\def\treidta{ \tau_{\rm R} (\mmk;\ta) }
\def\treidxo#1{ \tau_{\rm R} (#1;e^{i\tx}) }
\def\kth{ {\rm K3} }
\def\teta{ \tilde{\eta} }
\def\tchi{ \tilde{\chi} }
\def\ito{ I_{\thetx,1} }
\def\itt{ I_{\thetx,2} }

\def\invor#1{ {#1}^{\ast} }

\def\chif#1{ \omega^{(#1)} }
\def\chift#1{ \omega^{(#1)} }
\def\chivar{ \chi }

\def\btor{ x\in \partial(M\setminus \cK) }
\def\mmk{ M\setminus \cK }

\def\ketpsi#1{ |\psi^{(#1)} \rangle }

\def\ta{ t }

\def\Al{ \Delta_{\rm A}(\cK;\ta) }
\def\Alder{ \Delta_{\rm A}^{\prime\prime}(\cK) }
\def\Aldersh{ \lrbc{\Alder - {1\over 12} } }

\def\inter{ L }
\def\Vbun{ V }
\def\Vfib{ V_{\cvphi} }

\def\hsg{ \cH_{\Sigma_g} }
\def\fp{ (-1)^F }

\begin{document}

\begin{titlepage}
\vspace*{-1.2in}
\centerline{\hfill                 IASSNS-HEP-96/128, hep-th/9612216}
\begin{center}

{\large \bf
Hyper-K\"{a}hler Geometry and Invariants of Three-Manifolds
}
\\
\bigskip
\bigskip
\bigskip

\centerline{L. Rozansky
}

\centerline{\em Department of Mathematics}
\centerline{\em University of Illinois at Chicago}
\centerline{\em Chicago, IL 60607, U.S.A.}
\centerline{{\em E-mail address: rozansky@math.uic.edu}}

\bigskip

\centerline{E. Witten
}

\centerline{\em School of Natural Sciences}
\centerline{\em Institute for Advanced Study}
\centerline{\em Princeton, NJ 08540, U.S.A.}
\centerline{{\em E-mail address: witten@sns.ias.edu}}

\bigskip

\vfill
\bigskip
\bigskip

\end{center}
\begin{quotation}
\noindent
{\bf Abstract.} We study a 3-dimensional topological sigma-model,
whose target space is a \hk manifold $\hx$. A Feynman diagram
calculation of its partition function demonstrates that it is a
finite type invariant of 3-manifolds which is similar in structure to
those appearing in the perturbative calculation of the Chern-Simons
partition function.

The sigma-model suggests a new system of weights for finite type invariants
of 3-manifolds, described by trivalent graphs. The Riemann curvature of $\hx$
plays the role of Lie algebra structure constants in Chern-Simons
theory, and the Bianchi identity
plays the role of the Jacobi identity in guaranteeing the so-called IHX
relation
among the weights.

We argue that, for  special choices of $\hx$,
the partition function of the sigma-model yields the Casson-Walker
invariant and its generalizations.  We also
derive Walker's surgery formula from the $SL(2,\ZZ)$ action on the
finite-dimensional Hilbert space obtained by quantizing the sigma-model
on a two-dimensional torus.

\end{quotation}
\vfill
\end{titlepage}

\pagebreak

\tableofcontents
\nsection{Introduction}
\label{s0}

The invariants of 3-dimensional topology which are based on quantum
Chern-Simons theory\cx{Wi1} contain a wealth of information, in some ways
almost too much.
One would like to be able to extract simple statements and in particular
to compare these invariants to more classical invariants of topology.
For this one might want something more elementary than the full-fledged
Chern-Simons theory.

A possible approach to study the structure of the quantum invariants is
to look at their semi-classical asymptotics. Let us recall that from
the quantum field theory point of view,
one constructs quantum invariants of
three-manifolds with the following starting point.  One introduces a
gauge group $G$, usually compact, and one considers a connection $A$
on a $G$-bundle $E$ over an oriented  three-manifold $M$, with
the Chern-Simons action
\qq
\Scs={1\over 2}\int_M\Tr \left(A\wedge dA+{2\over 3}A\wedge A
\wedge A\right).
\label{lagr}
\qqq
Here $\Tr $ is a quadratic form on the $G$ Lie algebra with certain
integrality properties.
The quantum invariant of $M$ (also known as the partition function) is
expressed as a path integral over (the gauge equivalence classes of)
connections $A$
\qq
\zcs = \int \exp\lrbc{ {ik\over 2\pi} \Scs } {\cal D} A;
\label{pf}
\qqq
here $k$ is an integer.
As the definition of $\Scs$
requires no metric on $M$, the partition function\rx{pf}
 is formally expected to be a topological invariant of $M$.
Closer investigation shows (at a physical level of rigor) that this is
actually so, modulo some subtleties about a framing of $M$ that must be
introduced in the quantization.

The form of the integrand in \ex{pf} suggests that the behavior
of $\zcs$ in the semi-classical regime (\ie near $k=\infty$) should
be governed by a stationary phase expansion. The starting point
of such an expansion is to pick a critical point
$A_0$ of $\Scs$, that is to say, a flat connection on $E$.  Then
one expands the quantum theory about $A=A_0$, generating Feynman
diagrams in a standard fashion\cx{AxSi},\cx{AxSi1},\cx{BN1},\cx{Ko}.
(The analysis is most simple if the flat connection $A_0$ is isolated
modulo gauge transformations; otherwise one meets some additional
complications, analyzed in this context in\cx{AxSi2}.)
The asymptotic expansion around the $c^{th}$  flat connection (or more
generally,
the $c^{th}$ component of the moduli space of flat connections)
produces an asymptotic series $\zcsc$ in powers of $1/k$.
The exact Chern-Simons invariant $\zcs$ (which is defined as an actual
function of $k$, at least for positive integer arguments, in contrast
to the individual $\zcsc$ which may very possibly only exist in general
as asymptotic series) has to all orders in $1/k$ an asymptotic expansion of
the form
\qq
\zcs = \sum_c \zcsc.
\label{sum}
\qqq
Each  of the $\zcsc$ is  the  product of the ``classical exponential''
$\exp\lrbc{ {ik\over 2\pi}\Scsc }$ (here $\Scsc$ is a Chern-Simons
invariant of a classical flat connection in the $c^{th}$ component) multiplied
by some determinants and by an asymptotic
series in $k^{-1}$.  In this
expansion,
the term of relative order $k^{-n}$ in the asymptotic
series is expressed in terms of trivalent graphs with $n+1$
loops. The graphs are trivalent because $\Scs$ is a cubic function of
$A$.

If the flat connection $A_0$ is non-trivial, one really constructs
in this way an invariant of the three-manifold $M$ together with a
representation of its fundamental group in $G$ (which determines the
flat connection).  If one wants to
get a ``pure'' three-manifold invariant, one approach is to take
$A_0$ to be the trivial flat connection and consider its contribution
$\ztrcs$ in the sum\rx{sum}.
One finds that the $G$-dependence of the three-manifold invariant
$\ztrcs$ can be described
very simply, provided that $M$ is a rational homology sphere, which ensures
that the trivial flat
connection is isolated.  (Otherwise, the trivial flat connection is a point
on a moduli space of flat connections over which one must integrate, and
things become more complicated.)  In fact, many of the ingredients
are independent of $G$.
To each $n+1$-loop trivalent  graph $\Gamma$, one associates
a certain fairly complicated integral $I_\Gamma$
of a product of Green's functions
over a product $M\times M\times \dots \times M$.  $I_\Gd(M)$ depends
on a metric on $M$ (or some other ``gauge fixing'' data that break the
topological symmetry), but not on $G$.  One also associates to
each such graph a ``weight factor'' $a_{\Gd}(G)$ which depends on
$G$ but not on $M$. The trivial connection contribution is
proportional to the (exponential) of an asymptotic series
\qq
\exp\lrbc{ \sum_{n=1}^\infty S_{n+1,G}(M) k^{-n}}
\label{expsum}
\qqq
The coefficient $S_{n+1,G}(M)$ in the asymptotic
expansion of $\ztrcs$
for gauge group
$G$ is then
\qq
S_{n+1,G}(M)=\sum_{\Gd\in\Gs_{n,3}} a_{\Gd}(G) I_\Gd(M),
\label{sumint}
\qqq
where the
sum runs over all trivalent graphs $\Gd$ with $n+1$ loops (and
$2n$ vertices). The Jacobi identity of the Lie algebra of $G$ is
used to show that although the individual integrals $I_\Gd(M)$
depend on the metric, the metric-dependence cancels out of the sum.

Part of the interest in the coefficients $S_{n,G}(M)$
is related to the fact that they are
expected to be related in a relatively accessible  fashion to classical
invariants, but with the information organized in a new way suggested
by quantum field theory.   Indeed, the coefficient in front of the
exponential\rx{expsum} is known to be related
to the order of the first homology group $\ordH$\cx{FG}
and the second coefficient is proportional to the Casson invariant of
$M$\cx{RoS1},\cx{Ro1}\footnote{In the case of Seifert rational
homology spheres, the relation between the Casson invariant and
two-loop perturbative correction was observed independently by
J.~Andersen by analyzing some results of\cx{RoS1}. }.

The asymptotic coefficients $S_{n,G}(M)$
 of $\ztrcs$ fall into the category of the
so-called ``finite type'' invariants of 3-manifolds $M$, which were
introduced in\cx{Oh1} (see\cx{BN2} for an exhaustive review of the
properties of finite type invariants of knots). Loosely speaking,
these invariants can detect the features of $M$ which are only of
limited (\ie finite) complexity. The coefficient $S_{n,G}(M)$ is
finite type of order $3(n-1)$.

Let $\Gs_{n,3}$ be the set of all closed graphs with $2n$ trivalent
vertices. We assume that there is a cyclic ordering of legs at each
vertex. A function on $\Gs_{n,3}$ is called a weight
if it is antisymmetric under the permutation of legs at a vertex and
if it satisfies the so-called IHX relation (see \eg\cx{BN2} for its
definition). It was shown in\cx{LMO} that for every weight function
one can construct a finite type invariant of order $3n$. The weight
functions of the coefficients of $\ztrcs$ are calculated by placing
the structure constants of the Lie group of $G$ at the vertices of
trivalent graphs and contracting their indices along the edges with
the help of quadratic form $\Tr$. The IHX relation for these
weights follows from the Jacobi identity.

There has been considerable interest in the question of whether there
are other weight factors $b_\Gamma$, not derived from Lie groups, such
that $\sum_\Gamma b_\Gamma I_\Gamma$ is likewise an invariant of
rational homology spheres.
There has also been much interest in relating these graphical
invariants to three-manifold invariants constructed in other ways.

In the present paper, we will obtain two results bearing on these
questions:
\begin{itemize}
\item[(I)]
For any compact or asymptotically flat \hk manifold $\hx$ of
real dimension $4n$, we construct weight functions $b_\Gd(\hx)$ (see
\ex{3.p3}) on closed
trivalent graphs with up to $2n$ vertices. We also construct weight
functions on trivalent graphs which have external legs. We provide a
simple mathematically rigorous proof that the weights satisfy the
IHX relation, thus demonstrating that the linear combinations of
integrals $\sum_{\Gd\in\Gs_{n,3}} b_\Gd(\hx) I_\Gd(M)$
are topological
invariants of rational
homology spheres $M$. As in the Chern-Simons case, there is also a
corresponding statement for general oriented 3-manifolds.\footnote{There
is a difference, though: in the Chern-Simons case increasing $b_1$ makes
the analysis more complicated, while here it becomes more simple.}
\item[(II)]
We show that the (unique) invariant that can be
constructed from the integrals $I_\thetx(M)$ associated with
the two-loop graph (we call this graph $\thetx$) is proportional to
the $SU(2)$ Casson-Walker invariant. Although we do not obtain
precise formulas, our construction shows that the analog of
Casson-Walker invariant based on a Lie group of rank $r>1$\cx{LM} is
a linear combination of integrals $I_\Gd(M)$ related to trivalent
graphs with $r+1$ loops.
\end{itemize}

Each of these two results can be understood, to a certain extent,
without the machinery we will develop.  In particular, once one
suspects that it is true, (I) can be checked by a direct calculation.
Similarly, the result (II) is not new, having been obtained
before\cx{RoS1},\cx{Ro1} by analyzing the
surgery formula for the partition function of $SU(2)$ Chern-Simons
theory.
However we hope to show that these results have a very natural origin
in a certain quantum field theory context.

The natural context is in fact a new type of 3-dimensional topological field
theory.
This theory is a twisted version of an $N=4$ supersymmetric
sigma-model. The sigma-model partition function $Z_\hx(M)$ is
expressed as a path integral over the maps from $M$ into a (compact
or asymptotically flat) \hk manifold $\hx$. For a $4n$-dimensional
\hk manifold $\hx$, a perturbative calculation of $Z_\hx(M)$ presents
it as a sum over closed graphs with $2n$ $(3-b_1(M))$-valent
vertices
\qq
Z_\hx(M) = \sum_{\Gd\in\Gs_{n,3-b_1(M)} }b_\Gd(\hx) I_\Gd(M).
\label{signs}
\qqq
Here $b_1(M) = \dim H_1(M,\IR)$ is the first Betti number of $M$. For
$M$ with $b_1(M)=0$, $b_\Gd(\hx)$ present new weight functions on
trivalent graphs. Instead of Lie algebras, they are based on compact
or asymptotically flat \hk manifolds. Roughly speaking, the Riemann
curvature of $\hx$ plays the role of Lie algebra structure constants
in writing the expression for $b_\Gd(\hx)$, and the Bianchi identity
plays the role of the Jacobi identity in the proof of the IHX
relation. This explains our first claim.

The claim (II) has a perhaps more subtle origin than (I).
$N=2$ supersymmetric gauge theory in four dimensions with
gauge group $SU(2)$ can be topologically
twisted\cx{Wi2} to get a theory whose correlation functions are
Donaldson invariants of  four-manifolds. If dimensionally reduced to
three dimensions, this theory has a twisted version with the
Casson-Walker invariant for its partition function (see, \eg\cx{AJ}
for a review). On the other hand, this three-dimensional theory
reduces at low energies\cx{SeWi2} to a
sigma-model whose target space is a certain smooth, non-compact but
asymptotically flat four-dimensional \hk manifold $\ah$, which in
fact coincides with the two-monopole moduli space whose geometry was
described in detail in\cx{AH}. So the Casson-Walker
invariant coincides with the invariant computed from the
sigma-model with target space $\ah$; this invariant is a multiple
of the unique two-loop graphical invariant, explaining our second
claim.

If $SU(2)$ is replaced by a gauge group $G$ of  rank $r$, the
two-monopole moduli space would simply be replaced by the moduli
space of vacua of the three-dimensional supersymmetric gauge theory
(with $N=4$ supersymmetry in three dimensions, corresponding to $N=2$
in four dimensions) with gauge group $G$. This moduli space is of
dimension $4r$, so the Casson-Walker invariant of $G$ for rational
homology spheres should be a finite type invariant of order
$3r$ which is computable from trivalent graphs with $r+1$ loops (but
we do not know precisely which weights are required).
In fact, it has recently been argued that the moduli space for $G=SU(n)$
is simply the reduced moduli space of BPS $n$-monopoles of $SU(2)$
\cx{chalmers},\cx{hanany}.  This fact could be used, in principle,
to obtain detailed formulas for $SU(n)$.

In Section 4, we will compare in detail the invariant we obtain
for the case where the target is $\ah$ to the Casson-Walker
invariant, as extended to arbitrary three-manifolds by Lescop
\cx{Lescop}.  The details depend very much on the value of the first
Betti number $b_1(M)$.
We find that the invariant $Z_\ah(M)$ computed
from the sigma-model and the Casson-Walker-Lescop invariant agree up
to a sign factor
\qq
Z_\ah(M) = {1\over 2} (-1)^{b_1} \lc(M).
\label{ahl}
\qqq
Here $\lc(M)$ is {\em twice} the Casson invariant as defined in\cx{Lescop}
(roughly speaking, in our normalization $\lc(M)$ measures the Euler
characteristic of the moduli space of flat $SU(2)$ connections on $M$).

For compact $X$, we expect that the theory under consideration in this
paper will obey the full axioms of a (non-unitary)
topological quantum field theory  as formalized by Atiyah
\cx{topax}.  (Unitarity would mean that the vector spaces
${\cal H}_\Sigma$ associated to two-manifolds $\Sigma$ have hermitian
metrics compatible in a natural way with the rest of the data; that is
so in Chern-Simons theory but not in the theory considered here.)
Proving this  goes beyond the scope of the present paper, though we develop
many of the relevant facts in Section 5.
We suspect that a full direct proof would be far
simpler than the corresponding analysis of
Chern-Simons theory\cx{sawin}.
If $X$ is not compact, one cannot quite get this structure, as we will see
in Section\rw{s6}.

In the analysis, we will meet two  ``anomalies'' which are
``curable.'' The path integral representation of the partition function
$Z_\hx(M)$ requires that $M$ be equipped with a 2-framing and an
orientation on the sum of (co)homology spaces
\qq
2H^0(M,\IR)\oplus 2H^1(M,\IR);
\label{hom}
\qqq
here $2H$ denotes $H\oplus H$.
However, since there is a canonical choice of 2-framing\cx{At} and
a canonical choice of orientation for the space\rx{hom},
the partition function $Z_\hx(M)$
can be transformed into a genuine invariant of an oriented
3-manifold.

The problem treated in the present paper is the three-dimensional analog
of ``integrating over the $u$-plane'' in Donaldson theory.  Some of the
features we find should have analogs for four-manifolds with small values
of $b_1$ and $b_2^+$.

\nsection{The topological sigma-model}
\subsection{Review of \hk geometry}
\label{s1}

We begin with a review of those few aspects of \hk geometry which
we will need.

A \hk manifold $\hx$ is a manifold of real
dimension $4n$
\qq
\dim_\IR \hx = 4n
\qqq
($4n$ will denote the dimension of $\hx$ throughout the paper). $\hx$
has
a Riemannian metric such that the holonomy of the Levi-Civita connection
lies in an $Sp(n)$ subgroup of $SO(4n)$.  This means that the complexification
of the tangent
bundle $T\hx$ of $\hx$ decomposes as $T\hx\otimes_{\bf R}{\bf C}
=V\otimes S$, where $V$ is a rank
$2n$ complex vector bundle with structure group $Sp(n)$, and $S$ is
a trivial rank two bundle.  In fact, $SO(4n)$ contains a subgroup
$(Sp(n)\times SU(2))/\Z_2$; $Sp(n)$ acts on $V$ and $SU(2)$ (which equals
$Sp(1)$) acts on $S$.  The Levi-Civita connection on $T\hx_\C$ simply
reduces to a $Sp(n)$ connection on $V$ (times the trivial connection on $S$).
The bundle $V$ will appear throughout the paper.

Sometimes, we will pick on $\hx$ local coordinates $\phi^i$; the metric
of $\hx$ will be called $g_{ij}$.
The indices $A,B,\ldots =1,2$ will label the two-dimensional representation
of $Sp(1)$,  while the indices
$I,J,\ldots =1,\dots, 2n$
will refer to the $2n$-dimensional representation of
$Sp(n)$.

The trivial $SU(2)$ bundle $S$ is endowed with an $SU(2)$-invariant
antisymmetric tensor $\epsilon_{AB}$. The antisymmetric
tensor $\eps^{AB}$ is defined as the inverse of $\eps_{AB}$
\qq
\epsilon^{AB} \eps_{BC} = \delta^A_C.
\label{1.1}
\qqq
Similarly, the fact that $V$ has structure group $Sp(n)$ means that
there is an invariant antisymmetric tensor $\epsilon_{IJ}$ in $\Wedge^2V$.
We write its inverse as $\epsilon^{IJ}$, defined so that
\qq
\epsilon^{IJ}\epsilon_{JK}=\delta^I_K.
\label{fun}
\qqq
The $\epsilon$ tensors are used to raise and lower $Sp(1)$
and $Sp(n)$ indices (it will hopefully cause no confusion to refer
to both of these invariant tensors as $\epsilon$).
Finally, the  decomposition $T\hx_C=V\otimes S$
corresponds to the existence of covariantly constant tensors
$\gamma_i^{AI}$ and $\gamma^i_{AI}$ that describe the maps from
$V\otimes S$ to $T\hx$ and vice-versa.

Now let us describe the form of the Riemann tensor on a \hk
manifold.  In general, the Riemann tensor is a two-form
$R_{ij}$ ($=-R_{ji}$) valued in the Lie algebra of $SO(4n)$.  The
condition of $Sp(n)$ holonomy means that $R$ actually takes values
in the Lie algebra of $Sp(n)$.  This is equivalent to the statement
that $R_{ij}=\gamma_i^{AI}\gamma_j^{BJ}\epsilon_{AB}R_{IJ}$,
where $R_{IJ}=R_{JI}$ is a two-form with values in ${\rm Sym}^2\,V$.
(Recall that the Lie algebra of $Sp(n)$ can
be identified with the {\it symmetric} matrices $M_{IJ}=M_{JI}$.)
Now let us write explicitly
$R_{ij}= {1\over 2}R_{ij kl}dx^k\wedge dx^l$.
The symmetry of the Riemann
tensor with $R_{klij}=R_{ijkl}$ means that we can make the same
manipulation on the first two indices that we just made on the last
two, learning that
\qq
R_{ijkl}=-\gamma_i^{IA}\gamma_j^{JB}\gamma_k^{KC}\gamma_l^{LD}
\epsilon_{AB}\epsilon_{CD}\Omega_{IJKL},
\label{rform}
\qqq
for some $\Omega\in {\rm Sym}^2\,V\otimes {\rm Sym}^2\,V$ (that is,
$\Omega_{IJKL}=\Omega_{JIKL}=\Omega_{IJLK}$).
Finally, the property of the Riemann tensor
that $R_{ijkl}+R_{iklj}+R_{iljk}=0$ implies that $\Omega$ is
completely symmetric in all indices.

\subsection{Bianchi identity}

Now, armed with this knowledge of the structure of the Riemann tensor,
let us look at the Bianchi identity, which in general reads $DR_{ij}=0$
($D$ being
the exterior derivative defined using the Levi-Civita connection).
It reduces in the \hk case to
\qq
D_{AI}\Omega_{JKLM}=D_{AJ}\Omega_{IKLM},
\label{bid}
\qqq
where $D_{AI}=\gamma^i_{AI}D_i$.

Now, consider the second derivative of $\Omega$, antisymmetrized on
the $SU(2)$ and $Sp(n)$ indices:
\qq
\epsilon^{AB}\left(D_{AI}D_{BJ}-D_{AJ}D_{BI}\right)\Omega_{KLMN}.
\label{bite}
\qqq
To analyze this expression, we use first\rx{bid} to write
$D_{AI}D_{BJ}\Omega_{KLMN}=D_{AI}D_{BK}\Omega_{JLMN}$.
Since for any tensor $X_P$
\qq
\lrbs{ D_{AI}, D_{BK} }X_P =
\eps_{AB}\, \eps^{ST}\, \Omega_{PIKS}\, X_T,
\label{2.bit1}
\qqq
we get, after commuting the derivatives $D_{AI}D_{BK}$,
\qq
D_{AI}D_{BK} \Omega_{JLMN} & = &
D_{BK}D_{AI} \Omega_{JLMN} +
\eps_{AB}\, \eps^{ST} \Omega_{JIKS} \Omega_{LMNT}
\label{2.bit2}\\
&&\qquad +
\eps_{AB}\, \eps^{ST} \lrbc{ \jacbi }.
\nonumber
\qqq
Finally, using \ex{bid} again as
$D_{BK}D_{AI}\Omega_{JLMN} = D_{BK}D_{AL}\Omega_{IJMN}$, we see that
\qq
\eps^{AB} D_{AI} D_{BJ} \Omega_{KLMN} & = &
\eps^{AB} D_{BK} D_{AL} \Omega_{IJMN}
- 2 \eps^{ST} \Omega_{JIKS} \Omega_{LMNT}
\label{2.bit3}\\
&&\qquad -2 \eps^{ST} \lrbc{ \jacbi}.
\nonumber\\
\qqq
The \lhs of \ex{bite} is antisymmetric in $I,J$, while the first two
terms in the \rhs of \ex{2.bit3} are symmetric and the other terms
are antisymmetric in these indices. Therefore
\qq
\lefteqn{
\eps^{AB} \lrbc{ D_{AI}D_{BJ} - D_{AJ}D_{BI} } \Omega_{KLMN} }
\label{goodid}\\
&&\hspace*{1.5in}=
-4 \eps^{ST}\lrbc{\jacbi}.
\nonumber
\qqq

In verifying explicitly at the end of Section\rw{s3}
that certain Feynman diagram expressions
give three-manifold invariants,\rx{goodid} will play roughly the role
that the Jacobi identity for a Lie algebra plays in the analogous
calculation in Chern-Simons gauge theory.  A Lie algebra
${\cal J}$ with invariant quadratic form $\Tr$ is defined by a tensor
$f\in \Lambda^3{\cal J}$ obeying a Jacobi identity that is
schematically $ff+ff+ff=0$ (where indices are arranged differently in
the three terms). In replacing a Lie algebra by a \hk manifold, $f$
is replaced by the tensor $\Omega$, and the Bianchi identity by
\ex{goodid}, which reads schematically
$\Omega\Omega+\Omega\Omega+\Omega\Omega = {\rm total~ derivative}$.


\subsection{Background in six dimensions}
\label{s2}

\def\R{{\bf R}}
A supersymmetric sigma-model involving maps to a \hk manifold
$\hx$ can be defined first of all in six dimensions.
A superspace (or supermanifold) construction of this theory is not known,
but nevertheless the supersymmetric Lagrangian exists.
We describe this for background, though we will not actually work
in six dimensions in this paper.   In what follows, $\R^6$ is six-dimensional
Euclidean space with a flat metric.

The fields of the six-dimensional supersymmetric model
are a map $\Phi:\R^6\to \hx$ (which if we pick local coordinates $\phi^i$
on $\hx$ can be described by giving functions $\phi^i$ on $\R^6$) and
fermions $\psi$ taking values in ${\cal S}_+\otimes \Phi^*(V)$,
where ${\cal S}_+$ is one of the spin bundles of $\R^6$, and $V$ is the
$Sp(n)$ bundle over $\hx$ that entered in the last section.  Recall
that ${\cal S}_+$ is a rank four complex bundle.
In particular, $\Lambda^4{\cal S}_+$ is one-dimensional,
with a Lorentz-invariant generator which we will call $\epsilon$.
(This symbol will appear only in the present paragraph, and should cause
no confusion with the use of the same name for invariant antisymmetric
tensors on other bundles.)
If we use Greek letters $\alpha,\beta,\dots$ for ${\cal S}_+$-valued
objects, then $\epsilon$ is a
fourth rank antisymmetric tensor $\epsilon_{\alpha\beta\gamma\delta}$.
  The supersymmetric Lagrangian is
\qq
L=\int_{\R^6}d^6x\,\,\left({1\over 2}(d\Phi,d\Phi)
+{i\over 2}(\psi,D\psi)
+ {1\over 24}
\epsilon^{\alpha\beta\gamma\delta}\psi_{\alpha I}\psi_{\beta J}
\psi_{\gamma K}\psi_{\delta L}\Omega^{IJKL}\right).
\label{backg}
\qqq
Here $D$ is the Dirac operator mapping ${\cal S}_+\otimes \Phi^*V
\to {\cal S}_-\otimes \Phi^*V$, and $(~,~)$ are the natural metrics on
$T^*\R^6\otimes \Phi^*(T\hx)$ and ${\cal S}_-\otimes \Phi^*V$.

For our purposes, we want to reduce this theory to a dimension less than
six in which it can be ``twisted'' to give a topological field theory.
The largest such dimension is three.  Reducing the theory to three dimensions
means simply restricting to fields that are invariant under a
three-dimensional
group of translations of $\R^6$.  For instance, we pick coordinates
$x^1,x^2,\dots , x^6$ on $\R^6$, such that the metric is $\sum_i(dx^i)^2$,
and we require that the fields be independent of $x^4$, $x^5$, and $x^6$.
We are then left with a theory on $\R^3$ with an $SO(3)\times SO(3)$ symmetry,
where one $SO(3)$ -- acting on $x^1,x^2,x^3$ -- is the rotation symmetry
of the three-dimensional theory, and the other -- acting on $x^4,x^5,x^6$ --
is an ``internal'' symmetry.  We call these $SO(3)_E$ and $SO(3)_N$, and
we call their double covers $SU(2)_E$ and $SU(2)_N$.  Allowing for the
presence of the fermions, the symmetry group of the theory is actually
$(SU(2)_E\times SU(2)_N)/\Z_2$.

The supercharges of the theory transform as two copies of $({\bf 2},{\bf 2})$
under $SU(2)_E\times SU(2)_N$.  Now we ``twist'' the theory in the following
way.  We let $SU(2)'$ be a diagonal subgroup of $SU(2)_E\times SU(2)_N$,
and we define a new action of rotations by thinking of $SU(2)'$ as the
rotation generators.  The point of this is that the supercharges
transform as two copies of ${\bf 1}\oplus {\bf 3}$ under $SU(2)'$.
In particular, there are two $SU(2)'$-invariant supercharges.  If
we call these $Q_A$, $A=1,2$, then they obey $\{Q_A,Q_B\}=0$, as one
sees by restricting the underlying six-dimensional superalgebra.

Having rotation-invariant supercharges $Q_A$
that square to zero in the twisted theory usually means that the twisted theory can be
generalized from flat space to an arbitrary curved manifold (in this
case of dimension three) in a way that preserves conservation of the
$Q_A$ and such that the metric dependence is of the form
$\{Q_A,\dots\}$. When that can be done, then by restricting to the
$Q_A$-invariant observables, one gets a topological field theory.

In the case at hand, this program can be carried out.  The topological
field theory that one gets is written down in the next section.
Note that as the fermions of the untwisted theory transform
as $({\bf 2},{\bf 2})$ of $SU(2)_E\times SU(2)_N$, they transform
as ${\bf 1}\oplus {\bf 3}$ under $SU(2)'$.  In other words, the fermions
will be a zero-form $\eta$ and a one-form $\chi_\mu$.  Of course,
both $\eta$ and $\chi_\mu$ take values in $\Phi^*(V)$.  The formulas
in the next section were found by beginning with the conventional
$N=4$ supersymmetric sigma-model in three dimensions (which can be
obtained as explained above starting from the sigma-model in six
dimensions) and rewriting the standard formulas from the twisted
point of view. In the next section, we simply write down the result
without further commentary.

One important point is that non-linear sigma-models in three
dimensions are unrenormalizable and ill-behaved quantum mechanically,
and therefore we do {\it not} claim that the non-topological
observables of these theories are naturally defined.  In computing
the topological observables, it will turn out that everything comes
from low order terms that are not sensitive to the
unrenormalizability.  Therefore, pragmatically, we will not have to
commit ourselves to a particular point of view about this issue.  One
plausible point of view is that by adding higher derivative terms (of
the general form $\{Q,\cdot\}$) one could cure the
unrenormalizability (at some cost in beauty) without changing the
topological observables or our computations of them.


\subsection{The topological lagrangian and BRST symmetries}

We will now describe the topological sigma-model
that can be found using the recipe of the last subsection.
We work on an oriented three-dimensional
manifold $M$. We denote  local coordinates on $M$ as $x^\mu$,
$\mu=1,2,3$.  $M$ is endowed with a metric $h_{\mu\nu}$, but
an appropriate class of observables of the model will be metric-independent.
The target space of the sigma-model
is a $4n$-dimensional \hk manifold $\hx$.
The bosonic scalar fields are a map
$\Phi:M \rightarrow \hx$ which (once local coordinates are chosen on $\hx$)
can be described via functions $\phi^i(x^\mu)$, $i=1,\dots,4n$.
The fermions are a scalar $\eta^I$ and a one-form $\chi_\mu^I$ with
values in $V$.

The sigma-model action is
\qq
S & = & \int_M (L_1 + L_2) \sqrt{h} d^3 x,
\label{2.1}
\qqq
with
\qq
L_1 & = &
{1\over 2} g_{ij} \partial_\mu \phi^i \partial^\mu \phi^j +
\eps_{IJ} \chi^I_\mu \nabla^\mu \eta^J ,
\label{2.2}
\\
L_2 & = & {1\over 2} {1\over \sqrt{h} } \eps^{\mu\nu\rho}
\lrbc{\eps_{IJ}
\chi_\mu^I \nabla_\nu
\chi_\rho^J +
{1\over 3} \Omega_{IJKL} \chi_\mu^I \chi_\nu^J \chi_\rho^K \eta^L },
\label{2.3}
\qqq
The covariant derivative of fermions, here denoted as $\nabla_\mu$,
 is defined using the Levi-Civita
connection on $M$ and the pullback of the Levi-Civita connection on
$V$
\qq
\nabla_\mu = \partial_\mu\delta^I{}_J + (\partial_\mu \phi^i) \Gamma^I_{iJ}.
\qqq

The Lagrangians
$L_1$ and $L_2$ are each separately invariant under a pair of
BRST symmetries $Q_A$.  A transformation $\veps^A Q_A$
acts on the fields according to the formula
\qq
\delta_\veps \phi^i & = &
\gamma^i_{AI} \veps^A \eta^I,
\label{2.5}\\
\delta_\veps \eta^I & = & 0,
\label{2.6}\\
\delta_\veps \chi_\mu^I & = &
\veps^A \eps_{AB} \gamma_i^{BI} \partial_\mu \phi^i -
\Gamma^I_{iJ} (\delta_\veps \phi^i) \chi^J_\mu.
\label{2.7}
\qqq
The $Q_A$ obey
\qq
\{Q_A,Q_B\}=0,
\label{tobbey}
\qqq
as promised in the last subsection.

If among the infinitely many complex structures of the \hk
manifold $\hx$ (which are parametrized by a two-sphere) one picks
a particular one, and one lets $\phi^I$ be local holomorphic coordinates
in this complex structure, then one can pick a basis $Q,\overline Q$
for the two supercharges in which they act by
\qq
\begin{array}{lll}
\delta \phi^I = \eta^I, &\qquad &
\delta \phib^{\bI} =0, \\
\delta \eta^I = 0, &\qquad &
\delta \chi^I_\mu = \eps^{IJ} g_{J\bK} \partial_\mu \phib^{\bK}
- \Gamma^I_{JK} \eta^J \chi_\mu^K
\end{array}
\label{2.9}
\qqq
for $Q$ and
\qq
\begin{array}{lll}
\delta \phi^I = 0, &\qquad &
\delta \phib^{\bI} = g^{\bI J} \eps_{JK} \eta^K, \\
\delta \eta^I = 0, &\qquad &
\delta \chi_\mu^I = - \partial_\mu \phi^I
\end{array}
\label{2.10}
\qqq
for $\bQ$.

Now, let us verify formally that this theory is a topological field theory.
$L_2$ is manifestly independent of the metric of $M$; it is written
just in terms of wedge products and exterior derivatives of differential
forms.
$L_1$ is metric-dependent, but can be written as $\{Q_A,\cdot\}$
for any choice of $A$.  In fact,
\qq
\{\epsilon^AQ_A,\epsilon_{IJ}\gamma_i^{BI}\chi_\mu^J\partial^\mu\phi^i\}
=\epsilon^BL_1.
\label{longo}
\qqq
Since the metric-dependence of $S$ is thus of the form $\{Q,\cdot \}$,
the partition function and more generally the correlation functions
of $Q$-invariant operators are metric independent.

\subsection{Topological observables}

Now we want to introduce some topological observables. These are the
operators which (anti)commute with $Q_A$ but can not be presented as
(anti)commutators of $Q_A$ with some other operators. The vacuum
matrix elements of such operators are topological invariants of the
manifold $M$.

The simplest topological operators can be constructed from closed
forms on $\hx$. Once one picks a particular complex structure on $\hx$,
the tensor $\eps^{IJ} g_{J\bK}$ establishes an
isomorphism between the space $\Omega^{\ipe,0}$ of $(\ipe,0)$-forms
and the space $\Omega^{0,\ipe}$ of $(0,\ipe)$-forms on $\hx$. Let
$\omx$ be a $\ipe$-form which is $\partial$-closed as a
$(\ipe,0)$-form and $\bar{\partial}$-closed as a $(0,\ipe)$-form.
Then a similarity between the action\rx{2.9},\rx{2.10} of the BRST
operators $Q,\bQ$ on the pairs of fields $(\phi^I,\eta^I)$ and
$(\phib^{\bI},\eta^I)$, and the action of the operators
$\partial,\bar{\partial}$ on the differential forms on $\hx$ allows
us to conclude that the operator
\qq
\cO_\eta(\omx) = \omx_{I_1\ldots I_\ipe} \eta^{I_1} \cdots
\eta^{I_{\ipe}}
\label{2.on1}
\qqq
is BRST-closed.  These operators are further discussed in
Section\rw{s6}.

A more subtle topological operator can be constructed with the
help of the pullback of the connection $\Gamma^I_{iJ}$ from the $Sp(n)$
bundle $V$ over $\hx$ via the map $\Phi:\,M\rightarrow \hx$.
This pullback $\partial_\mu\phi^i\Gamma_i^I{}_J$ is not $Q$-invariant, and so
cannot be used to define knot invariants. However, it is possible to
modify it in order to achieve $Q$-invariance. Indeed, the modified
connection
\qq
A_{\mu IJ}  = \partial_\mu\phi^i \eps_{IK} \Gamma^K_{iJ} +
\Omega_{IJKL} \chi^K_\mu \eta^L
\label{2.o5}
\qqq
is readily seen to be $Q$-invariant up to a gauge transformation.
Hence if $\cK$ is a knot in $M$
and $\Va$ is a representation of $Sp(n)$, then the operator
\qq
\cO_\wa(\cK) = \Tr_\wa \Pexp \lrbc{ \oint_{\cK} A_{\mu IJ} dx^\mu },
\label{2.o6}
\qqq
which is the trace of the holonomy of $A_{\mu IJ}$ along $\cK$ taken
in the representation $\Va$, is $Q$-invariant and independent of the
metric on $M$. The expectation value of the product of such operators
should give invariants of framed links, rather as in Chern-Simons
gauge theory. For example, for knots in $S^3$, one would expect to
obtain knot invariants of finite type, generalizing those obtained
from the gauge theory.

\subsection{Sigma-model and Chern-Simons theory}

It may now be time to begin to
disclose the secret of how to think about this
theory.  $L_1$ and $L_2$ play a completely different role.
The fact that $L_2$ is metric-independent suggests that it is more
fundamental, so let us think about it first.

$L_2$ has a structure
very reminiscent of Chern-Simons theory, with the one-form $\chi$
playing the role of the connection $A$ of Chern-Simons theory.
Indeed, compare $L_2$, schematically
$\chi \wedge D\chi+\chi\wedge \chi\wedge \chi \eta$, to the
Lagrangian
of Chern-Simons theory, schematically
$A \wedge DA+A\wedge A\wedge A$ ($D$ is here the covariant derivative with
respect to a background flat connection about which we may be expanding).
It is true that in $L_2$ the interaction term is not just $\chi^3$ but
has an extra factor of $\eta$. However, as we will see, this does not
spoil the analogy\footnote{In fact, $\eta$ plays the role of a ``fermionic
coupling constant,'' making sure that we do not have to go too far in
perturbative expansions.}.

In quantizing Chern-Simons theory, one must introduce, in addition
to $A$, certain extra fields: anticommuting ghosts and antighosts
$c $ and $\bar c$  and
a commuting scalar auxiliary field $t$ (sometimes called the Nielsen-Lautrup
auxiliary field), all in the adjoint representation
of the group.\footnote{In
some gauges, $t$ can be eliminated while preserving locality.}
  This theory usually is quantized with just one fermionic
(BRST) symmetry, the transformation laws being
 $\delta \bar c=t,$ $\delta t=0$, and $\delta c^a=f^a_{bc}c^bc^c$,
$\delta A_\mu=-D_\mu c$,
with $f$ the structure constants of $G$.

Let us think of
our topological sigma-model as a sort of Chern-Simons theory with the
statistics reversed.  We make the following identifications:
$\eta$ corresponds to $t$, and after picking one of the complex
structures on $\hx$, $\bar c$ corresponds to the local antiholomorphic
coordinates $\bar\phi^{\bar I}$, and $c$ to the local holomorphic coordinates
$\phi^I$.  A peek back to the $\overline Q$ transformation laws
of the sigma-model shows that they match very nicely with the
BRST transformation laws of gauge-fixed Chern-Simons theory, as written
in the last paragraph.

Moreover, the analogy can be pursued further.  To quantize
Chern-Simons theory, one must add a gauge fixing term of the form
$\{Q,\cdot\}$.  Not only is $L_1$ likewise of the form $\{\bar Q,\cdot\}$,
but the match is much closer.  The most common gauge fixing term of
Chern-Simons theory is schematically $\Delta L=
\partial_\mu c\partial^\mu\bar c
+t D_\mu A^\mu$.  If we replace $t$ by $\eta$, $A$ by $\chi$,
$\bar \phi$ by $\bar c$, and $\phi$ by $c$,
$\Delta L$ does match up nicely with the detailed form of $L_1$.
The topological observable\rx{2.o6} corresponds to the holonomy
operator $\Tr\Pexp\lrbc{ \oint_{\cK} A }$ of the Chern-Simons theory
which produces the colored Jones polynomial. Note that the role of
the Chern-Simons gauge field $A$ is played in\rx{2.o6} by the
modifying term $\Omega\chi\eta$ of \ex{2.o5}. (We will later see that
the ``ordinary'' pullback term can be dropped.)

So our course is set.  We will try to compare the topological
sigma model to Chern-Simons theory, with the
curvature tensor $\Omega_{IJKL}$ of the \hk manifold playing
the role of the structure constants
$f_{abc}$ of the Chern-Simons gauge group.  Of course,
$f$ is antisymmetric while $\Omega$ is symmetric.  This is what comes
of reversing the statistics of the gauge field.  Moreover, $f_{abc}$ has
three indices while $\Omega_{IJKL}$ has four.  This actually will
lead to the main difference between the two theories.  The extra
index of $\Omega$ is coupled to an $\eta$ field, which will be absorbed
by an $\eta$ zero mode.  (Other contributions will be seen to vanish.)
As $\eta$ has a definite number of zero modes,
only diagrams with a definite number of vertices will contribute.
That is why the topological sigma-model will be related to three-manifold
invariants associated with trivalent graphs with a definite number of
loops -- in fact $n+1$ loops if $\hx$ has dimension $4n$.  Such a restriction
on the number of loops, which makes the theory vastly more elementary,
would be impossible in a unitary field theory
such as Chern-Simons theory, but the twisting that made the sigma-model
topological has spoiled unitarity.

Anticipating the analogy with Chern-Simons theory will greatly
facilitate the analysis of the next section.

\nsection{Perturbative calculations}
\label{s3}

Consider the partition function of the topological sigma-model
\qq
Z(M) = \pint \exp \lrbc{- \int (c_1 L_1 + c_2 L_2) \sqrt{h} d^3 x }.
\label{3.1}
\qqq
Here $L_{1,2}$ are the lagrangians \rx{2.2} and\rx{2.3}, and
$c_{1,2}$ are some arbitrary constants. Since $L_1$ is BRST-exact,
the partition function does not depend on $c_1$. A field rescaling
\qq
\eta^I \rightarrow \lambda \eta^I, \qquad
\chi^I_\mu \rightarrow \lambda^{-1} \chi^I_\mu
\label{3.2}
\qqq
does not change $c_1$ but changes $c_2$:
$c_2\rightarrow \lambda^{-2} c_2$.
Thus $Z(M)$ does not depend on
$c_2$ either.\footnote{This argument is valid
for Feynman diagrams of more than one loop, but not for the one-loop
determinants, which in any case will be examined closely below.}
We choose $c_1=c_2={1\over \hbar}$  so that
\qq
Z(M) = \pint \exp \lrbc{ - {1\over \hbar} S },
\label{3.3}
\qqq
where $S$ is the action\rx{2.1}. The partition function\rx{3.3} should
be independent of the value of $\hbar$.

For small $\hbar,$ some form of perturbative expansion should be valid.
The minima of the action are the constant maps of $M$ to $\hx$, so
we will expand around those.  One must as usual be careful with bosonic
and fermionic zero modes.  The bosonic zero modes are simply the constant
modes of $\phi$ -- displacements in the constant map of $M$ to $\hx$.
In expanding around a constant map of $M$ to $\hx$,
the fermionic zero modes, if $M$ is a rational homology sphere
(\ie if the first Betti number $b_1=0$), are
the constant modes of $\eta$; they are
equal in number to the number of components of $\eta$, which is
 $2n$ if $\hx$ is of dimension
$4n$.  If $M$ is not a rational homology sphere but has first Betti
number $b_1>0$, then there are in addition $2nb_1$ zero modes of $\chi_\mu$.

To take account of the bosonic zero modes in a perturbative expansion,
one must introduce ``collective
coordinates,'' and integrate over the space of all constant maps
of $M$ to $\hx$.  Thus,
we split the bosonic field $\phi^i(x)$ into a sum of a constant and
fluctuating part,
\qq
\phi^i(x) = \cvphi^i + \vphi^i(x),
\label{3.4}
\qqq
where $\cvphi^i$ is constant, and $\vphi^i(x)$ is required to be orthogonal
to the zero mode.
The zero modes of $\phi^i(x)$ are thus contained in the constant part
$\cvphi^i$.
We define a partition function $Z_\hx(M;\cvphi^i)$
of fixed $\cvphi^i$, and
obtain the partition function $Z_\hx(M)$
as an integral over $\hx$
\qq
Z_\hx(M) = {1\over (2\pi\hbar)^{2n} }
\int_\hx Z(M;\varphi^i_0) \sqrt{g}\, d^{4n} \cvphi^i.
\label{3.04}
\qqq

The perturbative calculation presents the integrand $\zphim$
of\rx{3.04} as a product of two factors
\qq
Z(M;\varphi^i_0) = \zz \zf.
\label{3.x04}
\qqq
Here $\zz$ is the 1-loop contribution of non-zero modes, while $\zf$
is the exponential of the sum of all Feynman diagrams of two or more
loops, in the background field of given $\varphi_0$.

\subsection{The one-loop contribution}

Let us first determine the one-loop contribution $\zz$. We work with
the part of the action\rx{2.1} which is quadratic in fluctuating
bosonic fields $\vphi^i(x)$ and in fermionic fields
$\eta^I,\chi^I_\mu$:
\qq
S_0 & = &
\int_M \sqrt{h}\, d^3 x \lrbc{{1\over 2}
g_{ij}\, \partial_\mu \vphi^i \partial^\mu \vphi^j +
\eps_{IJ} \chi^I_\mu \nabla^\mu \eta^J +
{1\over 2} {1\over \sqrt{h} }
\eps^{\mu\nu\rho} \eps_{IJ} \chi^I_\mu \nabla_\nu
\chi_\rho^J }
\label{3.5}
\qqq
(the tensors $g_{ij}$, $\eps_{IJ}$ and implicit Christoffel symbols
$\Gamma^i_{jk}$ in this formula are taken at the point $\cvphi^i$ of
$\hx$). The path integral of the   bosonic fields $\vphi^i$ gives a factor
\qq
\lrbc{ {\det}\p \Delta_{(0)} }^{-2n},
\label{3.05}
\qqq
where $\Delta_{(0)}$ is the Laplacian acting on zero-forms on $M$ and
${\det}\p$ means that we exclude the (constant) zero modes.

Now introduce an operator $L_-$ which acts on the direct sum of zero- and
one-forms on $M$:
\qq
L_-(\eta, \chi_\mu) =
(-\nabla^\mu \chi_\mu, \nabla_\mu \eta + h_{\mu\nu} {1\over \sqrt{h}}
\eps^{\nu\rho\lambda}\partial_\rho \chi_\lambda ).
\label{3.105}
\qqq
If we define a scalar product
\qq
\sc{\eta,\chi_\mu}{\eta\p,\chi_\mu\p} =
\int_M \sqrt{h}\,d^3x ( \eta\eta\p + h^{\mu\nu} \chi_\mu \chi_\nu\p),
\label{3.205}
\qqq
then the fermionic part of the action\rx{3.5} becomes a quadratic
form
\qq
{1\over 2} \eps_{IJ} \sco{\eta^I,\chi^I_\mu}{L_-}{\eta^J,\chi^J_\mu}.
\label{3.305}
\qqq
We assume that the lattice-regularized expression for the fermionic
integration measure of \ex{3.1} is
\qq
\cD \eta^I \cD \chi^I_\mu =
\prod_{x\in M} {\hbar^n\over n!}
\lrbc{\eps_{IJ}\,d\eta^I(x)\,d\eta^J(x)}^n
\prod_{x\in M} {\hbar^{3n}\over (3n)!} \lrbc{
\eps_{IJ}\, h^{\mu\nu} d\chi^I_\mu(x)\, d\chi^J_\nu(x)}^{3n},
\label{latreg}
\qqq
here $\prod_{x\in M}$ means a product over all nodes of the lattice
which approximates $M$. For this choice of integration measure,
the fermionic one-loop contribution, with zero modes removed, is
\qq
\lrbc{ {\det}\p L_- }^n,
\label{3.405}
\qqq
so that the total one-loop contribution of non-zero modes is
\qq
\zz = \lrbs{ { {\det}\p L_- \over \lrbc{ {\det}\p \Delta_0 }^2 } }^n.
\label{3.505}
\qqq
Note in particular that $Z_0(M;\varphi_0)$ in its form\rx{3.505}
is independent of $\varphi_0$, therefore we may denote it
simply as $Z_0(M)$.

The fermionic zero modes are the zero modes of the operator $L_-$.
More precisely, the zero modes of $\eta^I$ and $\chi^I_\mu$ are
harmonic 0- and 1-forms on $M$ tensored with the fiber $\Vfib$
of the $Sp(n)$
bundle $\Vbun\to \hx$  at the point $\cvphi$. Therefore,
as anticipated above,
\qq
\#(\mbox{zero modes of} \;\eta^I) & = & 2n,
\label{3.705}\\
\#(\mbox{zero modes of} \;\chi^I_\mu) & = & 2n\dh.
\label{3.805}
\qqq

Because of the fermion zero modes, a
product of $2n$ fields $\eta^I$ and $2n\dh$ fields $\chi^I_\mu$ has a
non-zero vacuum expectation value. The
products of
 $\eta^I$ and  $\chi^I_\mu$ zero modes are  elements of the spaces
\qq
H_\eta & = & \lmax\lrbc{ H^0(M,\IR)\otimes \Vfib },
\label{3.8a1}\\
H_\chi & = & \lmax\lrbc{ H^1(M,\IR)\otimes \Vfib}
\label{3.8a2}
\qqq
respectively.

The
absolute value of the
ratio
$\left|{ {\det}\p L_- / \lrbc{ {\det}\p \Delta_0 }^2 } \right|$
is equal to the analytic Ray-Singer torsion of (the trivial
connection on) $M$. Therefore its $n$th power $\zz$ is
not just a number but an element in the top exterior power of the space
$\lrbc{H_0(M,\IR)\oplus H^1(M,\IR)}\otimes \Vfib$, or, equivalently,
in the top exterior power of the space
\qq
\lrbc{ 2H_0(M,\IR)\oplus H^0(M,\IR)\oplus H^1(M,\IR)}\otimes \Vfib.
\label{3.8a3}
\qqq
Here $2H_0(M,\IR)$ denotes $H_0(M,\IR)\oplus H_0(M,\IR)$.
An element of the space
$$\lmax\lrbc{2H_0(M,\IR)\otimes \Vfib}$$
is a volume form for the integral
over the bosonic collective coordinates $\varphi_0$, which range over
$\hx$. Therefore $\zz$
determines the product of the volume form on $\hx$ and the vacuum
expectation values of the fermionic fields.

There are natural volume forms -- the Riemannian volume form
of $\hx$, and a measure for the fermions that will be written
presently.  $\zz$ is the product of the natural volume forms
times a number equal to $|H_1(M,\Z)|'$, the number
of points of finite order (\ie torsion elements)
in $H_1(M,\Z)$(see\cx{FG} and references therein).
Note that $|H_1(M,\Z)|'$ differs from the order of the first homology
$\ordH$, which is usually defined as follows:
\qq
\ordH =
\left\{
\begin{array}{cl}
\mbox{$\#$ elements in $H_1(M,\Z)$} &\mbox{if $\dh=0$}\\
0 &\mbox{if $b_1>0$}.
\end{array}
\right.
\label{3.h1}
\qqq
If $M$ is a rational homology sphere (\ie $b_1=0$) then, obviously,
$\ordH\p=\ordH$.

The natural volume forms (which must be multiplied by $|H_1(M,\Z)|'$
to get $Z_0(M)$) are as follows.
 There is a lattice inside
$H^1(M,\IR)$ which is formed by one-forms with integer integrals over
one-cycles of $M$. Let $\chift{\ia}_\mu$,
$1\leq \ia \leq \dh$ be
harmonic one-forms forming a basis of this lattice.
The natural measures for the fermion zero modes can be described by
giving
the fermionic vacuum expectation values
\qq
\langle \eta^{I_1}(x_1) \cdots \eta^{I_{2n}}(x_{2n}) \rangle
& = & \hbar^n \epsn^{I_1\ldots I_{2n} },
\label{3.8a4}
\\
\langle \chi^{I_1}_{\mu_1}(x_1) \cdots
\chi^{I_{2n\dh}}_{\mu_{2n\dh}} (x_{\dh}) \rangle
& = &
{\hbar^{n\dh}\over\lrbc{(2n)!}^{\dh}}
\sum_{s\in S_{2n\dh} }  (-1)^{|s|}
\label{3.8a5}\\
&&\hspace{-1.5in}\times
\prod_{\ia=0}^{\dh-1}
\left( \epsn^{I_{s(2\ia n+1)}\ldots I_{s(2\ia n+2n)} }
\chift{\ia}_{\mu_{s(2\ia n+1)}} (x_{s(2\ia n+1)} ) \cdots
\chift{\ia}_{\mu_{s(2\ia n+2n)}} (x_{s(2\ia n+2n)} ) \right)
\nonumber
\qqq
in calculating the Feynman diagrams which absorb the zero modes.
We used the following notation in \eex{3.8a4} and\rx{3.8a5}:
\qq
\epsn^{I_1\ldots I_{2n} } =
{1 \over (2n)!} \sum_{s\in S_{2n} }
(-1)^{|s|} \eps^{I_{s(1)} I_{s(2)} } \cdots
\eps^{I_{s(2n-1)} I_{s(2n)} },
\label{3.b5}
\qqq
$S_m$ is the
symmetric group of $m$ elements, and $|s|$ is the parity of a
permutation $s$.

A choice of an overall sign
in the formulas\rx{3.8a4} and\rx{3.8a5} for the fermionic expectation
values
is equivalent to a choice of orientations on the spaces
\qq
H^0(M,\IR)\otimes \Vfib, \; H^1(M,\IR)\otimes\Vfib.
\label{3.s1}
\qqq
As a result, the
sign of the whole partition function $Z_\hx(M)$ depends on this
choice. In other words, at a first glance, $Z_\hx(M)$ is an invariant
of $M$ with the choice of orientation on the spaces\rx{3.s1}, rather
than simply an invariant of $M$.
An orientation of the space $\Vfib$ is determined by the $n$th
power of the two-form $\eps_{IJ}$ on $\hx$. Since the space $\Vfib$
is even-dimensional, the orientation on the spaces\rx{3.s1} does not
depend on the choice of orientation on the cohomology spaces
$H^0(M,\IR),\;H^1(M,\IR)$ (this is why the sign of the expectation
value\rx{3.8a5}
does not depend on the choice of basic one-forms
$\chift{\ia}_\mu$). Therefore, the choice of orientation of the
spaces\rx{3.s1} and, consequently, the choice of the sign in
\eex{3.8a4},\rx{3.8a5}
(which was induced by the fermionic integration measure\rx{latreg})
is canonical and the invariant $Z_\hx(M)$ can
always be reduced to a canonical orientation.

This situation is similar
to the framing anomaly of the quantum Chern-Simons invariant\cx{Wi1}.
The Chern-Simons path integral is a topological invariant of framed
3-manifolds; however, since 3-manifolds have canonical
2-framings\cx{At}, it can be converted into a genuine invariant of
3-manifolds. Still, the framing anomaly shows up as a framing
correction in the surgery formula for the quantum Chern-Simons
invariant. We will see in subsection\rw{torus} that a similar sign
correction is required for the surgery formula of $Z_\hx(M)$.

\subsection{Minimal Feynman diagrams}

Now we will start to analyze
the Feynman diagrams that absorb the fermionic zero
modes. We may limit our attention to only those Feynman diagrams (we
call them ``minimal'') whose contribution is of order $\hbar^{2n}$.
Indeed, the integral\rx{3.04} related to $4n$ bosonic zero modes
carries a normalization factor $(2\pi\hbar)^{-2n}$. However, as we
have mentioned, due to the topological nature of the sigma-model,
the partition function $Z_\hx(M)$ should not depend on $\hbar$. Therefore
the contribution of all ``non-minimal'' Feynman diagrams cancels out.
Since there are always exactly $2n$  zero modes of $\eta$, the vanishing of
Feynman diagrams proportional to $\hbar^s$ with $s<2n$ will be fairly
obvious in what follows.  Less obvious in direct
inspection of the diagrams is the fact that the diagrams with
$s>2n$ also do not contribute.

We want to determine which  diagrams are minimal.
Consider a diagram with $V$ vertices.
Let $L$ be the total number of legs emanating from all of the various
vertices. Some
of the legs are joined together by propagators, others are attached
to fermionic zero modes. Each vertex in the diagram carries a factor
of $\hbar^{-1}$, so the vertices taken together
bring a factor of $\hbar^{-V}$. Each propagator
carries a factor of $\hbar$, and each fermionic zero mode carries a
normalization factor of $\hbar^{1\over 2}$. Therefore the total
contribution of propagators and external legs is $\hbar^{L\over 2}$.
So if the diagram is minimal, then
\qq
{L\over 2} - V = 2n.
\label{3.6}
\qqq

The  interaction vertices coming from the Lagrangians \rx{2.2}
and\rx{2.3} are -- in expanding around a constant map to $\hx$ --
of fourth  order and above, so
\qq
L\geq 4V.
\label{3.7}
\qqq
$\eta$ only appears linearly in the Lagrangian, so
all vertices are at most linear in $\eta^I$.
Therefore if a diagram contains
sufficient vertices to
absorb all $2n$ zero modes of $\eta^I$, then
\qq
V \geq 2n.
\label{3.8}
\qqq

The conditions\rx{3.6},\rx{3.7} and\rx{3.8} are compatible only if
$L=4V$ and $V=2n$.  We are thus reduced to a finite set of diagrams.
Moreover, the knowledge that
 the minimal diagrams should
contain only fourth order vertices which are proportional to
$\eta^I$ leads to a drastic simplification
in the effective Lagrangian.
 The action\rx{2.1} contains two vertices with the requisite properties
\qq
V_1 & = & {1\over 6} {1\over \sqrt{h} } \epsilon^{\mu\nu\rho}
\Omega_{IJKL} \chi^I_\mu \chi^J_\nu \chi^K_\rho \eta^L,
\label{3.9}\\
V_2 & = & -{1\over 2}\gamma_i^{AK}\gamma_j^{BL}\epsilon_{AB}
\Omega_{IJKL} \chi^I_\mu \eta^J
(\partial^\mu \vphi^i) \vphi^{j}.
\label{3.10}
\qqq
The vertex $V_1$ comes directly from $L_2$ of \ex{2.3}, while $V_2$
comes from the expansion of the connection $\Gamma^I_{iJ}$, which
appears in the covariant derivative $\nabla^\mu$ in the second term
of the \rhs of \ex{2.2}, in powers of $\vphi^i$.
Thus for the purpose of calculating the minimal diagrams in our
sigma-model, we can use the ``minimal'' Lagrangian
\qq
L_{\rm min} & = &
{1\over 2}g_{ij} \partial_\mu \vphi^i \partial^\mu \vphi^j +
\eps_{IJ} \chi^I_\mu \partial^\mu \eta^J +
{1\over 2}{1\over \sqrt{h} } \epsilon^{\mu\nu\rho} \eps_{IJ}
\chi^I_\mu \partial_\nu \chi^J_\rho
\nonumber\\
&&\qquad
+ {1\over 6} {1\over \sqrt{h} } \epsilon^{\mu\nu\rho}
\Omega_{IJKL} \chi_\mu^I \chi_\nu^J \chi_\rho^K \eta^L
-{1\over 2}\gamma_i^{AK}\gamma_j^{BL}\epsilon_{AB}
  \Omega_{IJKL} \chi^I_\mu \eta^J
(\partial^\mu \vphi^i) \vphi^{j}.
\label{3.11}
\qqq

The analogy -- described at the end of the last section --
between this theory and Chern-Simons theory can  now be
considerably perfected.
Instead of a non-polynomial sigma-model action, we now have reduced
the discussion to a polynomial action.  Moreover, though the vertices
are quartic, each vertex is linear in $\eta$.  Since we have $2n$ zero
modes of $\eta$ and will be looking at diagrams with precisely $2n$
vertices, $\eta$ will never appear in propagators; all factors of $\eta$
in vertices will go immediately into absorbing zero modes.  What is more,
as the zero mode wave functions are constant,
the absorption of the extra $\eta$ zero mode at each vertex will add no
additional complications to the integrals over $M$ associated with the
diagrams.  After thus eliminating
the $\eta$'s from the vertices, we reduce to a theory in which the vertices
are all cubic, just as in Chern-Simons theory.  Using the dictionary
comparing the sigma-model to Chern-Simons theory that was described
at the end of the last section, it can further be seen that the
vertices in the sigma-model have just the same structure as the vertices
(namely $A\wedge A\wedge A$ and $\bar c A_iD^ic$)
of gauge-fixed Chern-Simons theory.  The diagrams of this theory
will thus coincide with the diagrams of Chern-Simons theory, but the
weight factors are different, since the vertices given in the last
paragraph are proportional to the curvature tensor $\Omega$ of $\hx$,
rather than the structure constants of a Lie group.

\subsection{Propagators}

Each minimal diagram contains a total of $2n$ vertices\rx{3.9}
and\rx{3.10}. All fields $\eta$ in these vertices are used to absorb
the zero modes, so we need to know only the propagators
$\langle \vphi^i(x_1) \vphi^{j}(x_2) \rangle$ and
$\langle \chi^I_\mu(x_1) \chi^J_\nu(x_2) \rangle$. These propagators
can be expressed in terms of Green's functions $\Gphi$ and
$\Gchi{\mu\nu}$. $\Gphi$ is a symmetric function which satisfies the
equation
\qq
\Delta^{(1)} \Gphi = \delta(x_1 - x_2) - {1\over \Vol(M)}.
\label{3.13}
\qqq
Here $\Delta^{(1)}$ is a covariant Laplacian acting on the first
argument $x_1$, $\Vol(M)$ is the volume of $M$,
\qq
\Vol(M) = \int_M \sqrt{h}\, d^3 x,
\label{3.16}
\qqq
and the $\delta$-function is normalized by a condition
\qq
\int_M
\delta(x_1-x_2) f(x_2) \sqrt{h(x_2)}\, d^3 x_2
= f(x_1).
\label{3.17}
\qqq

$\Gchi{\mu\nu}$ is a symmetric one-form in both of its arguments.
It is co-closed  and has the following property: if $\cK_1,\cK_2$ are two
knots in $M$, then their linking number can be calculated as a double
integral
\qq
\oint_{\cK_1} dx_1^{\mu_1}
\oint_{\cK_2} dx_2^{\mu_2}
\,\Gchi{\mu_1\mu_2}.
\label{linking}
\qqq
A more formal way to define $\Gchi{\mu\nu}$ is to say that
there exists a function $\Gt{\mu\nu}$ which is a two-form in $x_1$
and a zero-form in $x_2$, such that
\qq
{1\over \sqrt{h(x_1)} } \epsilon^{\mu\nu\rho}
{\partial\over \partial x_1^\nu} \Gchi{\rho\lambda} =
\delta^\mu_\lambda \delta(x_1-x_2)
+ {1\over \sqrt{h(x_1)} } \epsilon^{\mu\nu\rho}
{\partial\over \partial x^\lambda_2} \Gt{\nu\rho}.
\label{3.14}
\qqq
$\Gchi{\mu\nu}$ also satisfies a condition
\qq
{\partial\over \partial x_1^\mu}\left( \sqrt{h(x_1)} h^{\mu\nu}(x_1)
\Gchi{\nu\rho}\right)=0.
\label{3.15}
\qqq

The propagators of $\vphi^i$ and $\chi^I_\mu$ are
\qq
\langle \vphi^i(x_1) \vphi^{j}(x_2) \rangle & = & - \hbar g^{ij}
\Gphi,
\label{3.18}
\\
\langle \chi^I_\mu(x_1) \chi^J_\nu(x_2) \rangle & = & \hbar \eps^{IJ}
\Gchi{\mu\nu}.
\label{3.19}
\qqq

\subsection{Feynman graphs and weight functions}
\label{s3w}

It is useful for the future discussion to introduce a notion of a
{\em Feynman graph} corresponding to a given Feynman diagram. The
Feynman graph is obtained by removing the legs of the Feynman diagram
which absorb the fermionic zero modes, and by ignoring the difference
between the $\vphi\vphi$ and $\chi\chi$ propagators.

The types of Feynman graphs participating in the calculation of
$\zf$ depend only on the dimension of the \hk manifold $\hx$ and on
the first Betti number $b_1$ of the
3-manifold $M$. $\hx$ determines the number of vertices in the
graphs, which is equal to $2n$.

If $M$ is a rational homology sphere (\ie if $b_1(M) = 0$) then there
are no $\chi$ zero modes to absorb. As a result, all $\chi$ fields in
the Feynman diagrams are connected by propagators and the
corresponding Feynman graphs are closed graphs with $2n$ tri-valent
vertices. The same type of graphs appeared in the calculation of the
trivial connection contribution to the quantum Chern-Simons invariant
of rational homology spheres.

If $b_1(M)=1$ then there are $2n$ zero modes of $\chi$ coming from a
harmonic one-form $\chif{1}_\mu$ on $M$. Since
$\chif{1}\Wedge \chif{1} = 0$, each vertex in a minimal Feynman
diagram can absorb at most one zero mode of $\chi$. Therefore each
vertex absorbs exactly one zero mode of $\chi$, and a
corresponding Feynman graph is a closed graph with $2n$ bi-valent
vertices. Obviously, such graphs are collections of circles with
vertex insertions. They appear also in the Chern-Simons calculation
of the Reidemeister torsion in the background of a $U(1)$ connection
proportional to the one-form $\chif{1}_\mu$.

If $b_1(M)=2$, then there are $4n$ zero modes of $\chi$ coming from
two harmonic one-forms $\chif{1},\chif{2}$ on $M$. Each vertex in a
minimal Feynman diagram can absorb at most one zero mode of $\chi$
proportional to a given one-form. Therefore, each vertex should
absorb exactly two zero modes of $\chi$, one coming from $\chif{1}$
and the other coming from $\chif{2}$. Since the vertex\rx{3.10} has
only one field $\chi$, it can not participate in the Feynman diagrams
when $b_1(M)>1$. The Feynman graphs have $2n$ uni-valent vertices, so
they look like a collection of $n$ segments. For reasons we will
explain later, the integral of a
product of a $\chi\chi$ propagator and one-forms $\chif{1},\chif{2}$
which corresponds to each segment, is proportional to a Massey product or
fourth order
``Milnor linking number'' of the forms $\chif{1}$ and $\chif{2}$.

If $b_1(M)=3$, then there are $6n$ zero modes of $\chi$ which are
proportional to three harmonic one-forms $\chif{1},\chif{2},\chif{3}$
on $M$. Each vertex in the Feynman diagrams absorbs three zero modes
of $\chi$ (one one-form of each kind). The Feynman graph is a collection
of $2n$ totally disconnected vertices with no edges.
Associated to each vertex is
an integral
$\int_M \chif{1}\Wedge\chif{2}\Wedge\chif{3}$ which measures the invariant
information that is contained in the cubic intersection form on $H^1(M,\Z)$.
For $b_1(M)=3$, the $M$-dependence of the partition function comes entirely
through this integral.

If $b_1(M)>3$, then no minimal Feynman diagram can absorb
all zero modes of $\chi$, and the whole partition function $Z_\hx(M)$
vanishes.

Denote the set of all closed graphs with $2n$
vertices, each of which is
$m$-valent,  as $\Gs_{n,m}$. Let $\zgmh$ be the  sum of contributions of
minimal Feynman diagrams which correspond to a graph $\Gd$. Then
the total contribution of Feynman diagrams is a sum over all Feynman
graphs permitted for a given pair of a \hk manifold $X$ and a
3-manifold $M$:
\qq
\zf = \sum_{\Gd \in \Gs_{n, 3-b_1(M)} } \zgmh.
\label{3.b1}
\qqq
Each contribution $\zgmh$ can be presented as a product of two factors
\qq
\zgmh = \wgh \sum_a \igi(M).
\label{3.b2}
\qqq
In this formula $\igi(M)$ are the integrals over $M$ of the products
of propagators $\Gphi$, $\Gchi{\mu\nu}$ as well as of the zero modes
$\chift{\ia}_\mu(x)$ coming from \ex{3.8a5}. The sum $\sum_a$ goes
over all possible ways to assign the vertices\rx{3.18} and\rx{3.19}
to the vertices of the Feynman graph $\Gd$. In other words, this sum
reflects the fact that for $b_1(M)=0,1$ different Feynman diagrams
may correspond to the same Feynman graph.

The factor $\wgh$ is a
product of tensors $\Omega_{IJKL}$ coming from the
vertices $V_1$ and $V_2$. Their indices are contracted by the
tensors $\eps^{IJ}$ contained in the propagators, and by the tensors
$\epsn^{I_1\ldots I_{2n} }$ contained in the zero mode expectation
values\rx{3.8a4} and\rx{3.8a5}. $\wgh$ also includes a sign factor
coming from rearranging the fermionic fields in the correlators.

The following technical paragraph contains
a precise prescription of how to calculate $\wgh$ for a
closed graph $\Gd$ with $2n$ trivalent vertices. We assign the
numbers $1,\ldots,2n$ to the vertices of the graph $\Gd$ and the
numbers $1,2,3$ to the legs of each vertex in the counterclockwise
order. We also assign the numbers $1,\ldots, 3n$ to the edges of the
graph and the numbers $1,2$ to the endpoints of each edge. Now the
graph $\Gd$ defines a map $\smg$ from the pairs $(k,l)$,
$1\leq k\leq 3n$, $l=1,2$ to the pairs $(i,j)$, $1\leq i\leq 2n$,
$j=1,2,3$ which describes how the edges are attached to the
legs of the vertices. Let $\sg$ be a permutation which maps a set of
pairs
$$
(1,0),(1,1),(1,2),(1,3),\ldots,(2n,0),(2n,1),(2n,2),(2n,3)
$$
onto a set
$$
(1,0),\ldots,(2n,0),\smg(1,1),\smg(1,2),\ldots,\smg(3n,1),\smg(3n,2).
$$
We denote by $|\sg|$ a parity of $\sg$. The function $\wgh$ is
defined as a product of tensors
\qq
\wgh =
(-1)^{|\sg|}\eps^{I_{1,0}\ldots I_{2n,0} }
\prod_{k=1}^{3n} \eps^{I_{\smg(k,1)}I_{\smg(k,2)}}
\prod_{l=1}^{2n}\Omega_{I_{l,0} I_{l,1} I_{l,2} I_{l,3}}.
\label{3.p3}
\qqq
Here we assume summation over repeated indices.

$\igi$ depends in an interesting way on $M$ but only rather trivially
on $\hx$: $\hx$ enters only through its dimension $4n$, which was seen
above to determine the numbers $V$ and $L$ of vertices and legs in our
diagrams.  Conversely, $\wgh$ depends on $M$ only through its first
Betti number, which determines how many legs will be used to absorb
the zero modes of $\chi$, and therefore which curvature integrals
must be performed on $\hx$. Combining \eex{3.04},\rx{3.6}
and\rx{3.b2} we find that
\qq
Z_\hx(M) = |H_1(M,\Z)|' \sum_{\Gd\in \Gs_{n, 3-b_1(M)} } \bg(\hx)
\sum_a \igi,
\label{3.b3}
\qqq
where
\qq
\bg(\hx) = {1\over (2\pi)^{2n} } \int_\hx \wgh \gdphi.
\label{3.b4}
\qqq

As we have already noted, when $M$ is a rational homology sphere,
the corresponding integrals $\igi$ are precisely the ones
that have already appeared in the
theory of perturbative Chern-Simons invariants. The ``perturbative''
sigma-model topological invariants appear as linear combinations
of these integrals, albeit with weights, given in \rx{3.b4}, that are
seemingly different than in the Chern-Simons case.  This is the basis
for claim (I) in the introduction.

We asserted in the introduction that once one suspects that invariants
of rational homology spheres can be constructed in this way, the invariance
can be checked directly, without reference to physics.  Let us see
how this comes about.  We assume some familiarity with the formal
proof of metric-independence of Chern-Simons perturbation theory,
as presented in\cx{AxSi},\cx{AxSi1}.

The basic idea is that when one varies with respect to the metric of $M$,
one of the propagators changes by an exact form.  After integration
by parts, this causes another propagator to collapse to a delta function,
giving a graph with one four-valent vertex in addition to the three-valent
vertices.  Each vertex has a factor of $\Omega_{IJKL}$ coupled to
various fields.  Upon collapsing a propagator, the $\Omega$ tensors
at the end are joined by an $\epsilon^{IJ}$ -- since the propagator
is $\epsilon^{IJ}$ times a form on $M\times M$.  The structure that
one gets is thus $\epsilon^{II'}\Omega_{IJKL}\Omega_{I'J'K'L'}$ with
the free indices coupled to various fields.  Most of the details
do not matter, but one important point is that as each interaction vertex is
linear in $\eta$, we have an $\eta$ zero mode contracted with one
index of each $\Omega$, so the structure is more precisely
$\epsilon^{II'}\Omega_{IJKL}\Omega_{I'J'K'L'}\eta^{L}\eta^{L'}$.  The
importance of this is that it means that we can assume antisymmetry
in $L$ and $L'$.

A given graph with one four-valent vertex and all other vertices trivalent
can arise in three different ways by collapsing a line or propagator
in a trivalent graph.  Just as in Chern-Simons theory, the three contributions
are identical except that the indices $JK$ and $J'K'$ become permuted,
so that the sum of the three contributions is proportional to
\qq
\epsilon^{II'}\left(\Omega_{IJKL}\Omega_{I'J'K'L'}+\Omega_{IJK'L}
\Omega_{I'J'KL'}+\Omega_{IJJ'L}\Omega_{I'KK'L'}\right)-
L\leftrightarrow L'.
\label{longo1}
\qqq

The vanishing of the expression\rx{longo1} as it appears inside the
integral\rx{3.b4} amounts to the IHX relation between the weights
of Feynman graphs. Therefore the following arguments constitute a
rigorous proof of the IHX relation for the weights $b_\Gd(\hx)$.

According to our previous identity\rx{goodid}, it is precisely the
expression\rx{longo1} that
can be written as
\qq
\epsilon^{AB}\left(D_{AL}D_{BL'}-D_{AL'}D_{BL}\right)\Omega_{JKJ'K'}.
\label{upongo}
\qqq
In Chern-Simons theory, instead of the quadratic function\rx{longo1}
of the curvature tensor, one obtains the quadratic
function of the structure constants that vanishes according to the
Jacobi identity.  Here, rather than zero, we get a sort of total derivative,
written in \rx{upongo}.  But in contrast to Chern-Simons theory, there is
an extra integral to do, namely the integral over $\hx$ in \ex{3.b4}.
The expression\rx{upongo}
does indeed disappear upon doing that integral.
Suppose, for instance, that we take the first term in \rx{upongo} and
integrate the left-most derivative $D_{AL}$ by parts.  Resulting
contributions would have $D_{AL}$ acting on the $\Omega$ tensor
at one or another vertex.  This gives an expression $D_{AL}\Omega_{STUV}$
with indices contracted with various fields.  Once again
many details are irrelevant.  The material point is that $L$ is contracted
with an $\eta$ field -- the antisymmetry in $L$ and $L'$ in\rx{longo1}
came because $L$ and $L'$ were contracted with $\eta$
fields -- and likewise (since all vertices in the minimal Lagrangian
are linear in $\eta$) one vertex in $\Omega_{STUV}$, say $V$, is
contracted with $\eta$. So we actually have something like
$\eta^L\eta^VD_{AL}\Omega_{STUV}$, and this vanishes because of the
Bianchi identity\rx{bid}.  (Of course, a contribution involving a
derivative of $\eta$ also vanishes because $\eta$ is to be contracted
with a covariantly constant zero mode.) This completes the direct
formal argument for the IHX relation of the weight
system\rx{3.b4} and for metric independence of the partition
function\rx{3.b3}.

At this point we would like to recall the following well-known fact
(see, \eg\cx{BN2}): the IHX relation implies that the weight
function vanishes on one-particle reducible graphs, \ie on
the connected graphs that can be made disconnected by removing one
edge. Therefore, since we have just established the IHX relation
for the weight system $\bg(\hx)$,
the one-particle reducible
diagrams can be excluded from
the list of minimal Feynman diagrams that we have
to consider in calculating the Feynman diagram contribution $\zf$.

By now
we have enough information about the structure of Feynman diagrams in
order to predict the behavior of the invariant $Z_\hx(M)$
under the change of orientation of $M$. Since the signs of the zero
mode expectation values\rx{3.8a4},\rx{3.8a5} do not depend on the
orientation of $M$, the only effect that its change has on $Z_\hx(M)$
comes from changing the sign in front of the tensor
$\epsilon^{\mu\nu\rho}$ and hence in front of the Lagrangian $L_2$ in
the action\rx{2.1}. As a result, the vertex $V_1$ and the
propagator $\chi\chi$ change signs.
Consider a minimal Feynman diagram containing $m$ vertices $V_1$ and
$2n-m$ vertices $V_2$. The number of $\chi\chi$ propagators in this
diagram is $n(1-b_1) + m$.
Therefore if the orientation of $M$ in changed,
then the invariant $Z_\hx(M)$ acquires a sign factor
\qq
(-1)^{n(1+b_1)}.
\label{orient}
\qqq

\subsection{Feynman diagrams with operator insertions}

Now let us see what happens when the operators\rx{2.on1}
are included in the path integral. The role of the operators $\cO_\eta$ of
equation\rx{2.on1}
is very simple. Each of them absorbs $\ipe$ zero modes of $\eta$, and
they do not interfere with Feynman diagrams. Therefore if we insert
operators $\cO_\eta$ with $m$ fields $\eta$, then a minimal Feynman
diagram should contain $2(n-m)$ vertices\rx{3.9},\rx{3.10} and
$n-m+1$ loops. In other words, the operators $\cO_\eta$ simply
reduce the ``effective'' dimension of the \hk manifold $\hx$.

In order to analyze the influence of the operators $\cO_\wa(\cK)$, we
expand the path-ordered exponential into a sum of cyclically ordered
integrals
\qq
\cO_\wa(\cK) = \Tr_\wa \sum_{k=0}^\infty {1\over k}
\eps^{J_1 I_2}\cdots \eps^{J_{k-1} I_k}
\intcycl A_{\mu_1 I_1 J_1}(x_1)\cdots A_{\mu_k I_k J_k} (x_k)\,
dx_1^{\mu_1} \cdots dx_k^{\mu_k}.
\label{2.o7}
\qqq
Here $\Tr_\wa$ refers to an uncontracted pair of indices $I_1,J_k$.
The analysis of diagrams with operators\rx{2.o7} is absolutely
similar to that for the operators $V_{1,2}$. Indeed, the operators
contained in the expansion of $A_{\mu I J}$ in fluctuating fields
$\vphi^i, \chi^I_\mu, \eta^I$ are at least quadratic, and at the
same time, they do not depend on $\hbar$. Therefore the minimal
Feynman diagram in the presence of operators $\cO_\wa(\cK)$ contains
only the ``modifying'' connection
\qq
B_{\mu IJ} = \Omega_{IJKL}\chi^K_\mu \eta^L
\label{2.o8}
\qqq
and the more obvious pullback term can be dropped.
We hence may use the simplified version of the
expansion\rx{2.o7}
\qq
\cO_\wa(\cK) = \Tr_\wa \sum_{k=0}^\infty {1\over k}
\eps^{J_1 I_2}\cdots \eps^{J_{k-1} I_k}
\intcycl B_{\mu_1 I_1 J_1}(x_1)\cdots B_{\mu_k I_k J_k} (x_k)
dx_1^{\mu_1} \cdots dx_k^{\mu_k}.
\label{2.o9}
\qqq

A minimal Feynman diagram contains altogether $2n$ vertices
$V_1,V_2,B_{\mu IJ}$. Therefore, if $\cL\subset M$ is an
$N$-component link with $Sp(n)$ representations $\Vab$ assigned to
its components $\cL_b$, $1\leq b\leq N$, then the partition function
\qq
\zamlh = \pint e^{-{1\over \hbar} S}
\prod_{b=1}^N \Tr_{\wa_b} \lrbc{ \Pexp \oint_{\cL_b}
A_{\mu IJ} dx^\mu }
\label{2.o10}
\qqq
can be expressed as a sum over Feynman graphs
\qq
\zamlh = |H_1(M,\Z)|' \sum_{\Gd\in \Gs_{n, 3-b_1(M),N} } \bg(\hx)
\sum_a \igi.
\label{2.o11}
\qqq
Here $\Gs_{n, 3-b_1(M),N}$ is a set of $\lrbc{3-b_1(M)}$-valent
graphs with legs attached to the components of the link $\cL$, while
$\igi$ are multiple integrals of the products of Green's functions
over $M$ and components of $\cL$, which have already appeared in
Chern-Simons perturbative calculations. The weights $b(\Gd)$ are
integrals\rx{3.b4}. The integrands $\wgh$ are calculated by the
formula which is similar to \ex{3.p3}.

To be more specific, consider a knot $\cK$ in a rational homology
sphere $M$. Let $\Gd$ be a graph with $2n-m$ trivalent vertices,
$3n-2m$ edges attached to trivalent vertices and $m$ legs attached to
the knot $\cK$.
A map $\smg$ from the set of pairs $(k,l)$, $1\leq k\leq 3n-2m$,
$l=1,2$ into the set of pairs $(i,j)$,
$1\leq i\leq (2n-m)$, $j=1,2,3$
still describes how the edges are attaced to the vertices of $\Gd$.
We assign numbers $1,\ldots,m$ to the legs in the
order in which they are attached to $\cK$. Then there
is a map
$\smg\p$ from $1,\ldots,m$ into a set of pairs $(i,j)$,
$1\leq i\leq 2n-m$, $j=1,2,3$, which shows how legs are
attached to trivalent vertices of $\Gd$. Let $\sg$ be a permutation
which maps a set of pairs
$$
(1,0),(1,1),(1,2),(1,3),\ldots,
(2n-m,3),
(1,0)\p,(1,1)\p,\ldots,(m,0)\p,(m,1)\p
$$
onto a set
\qq
&&(1,0),\ldots,(2n-m,0),(1,0)\p,\ldots,(m,0)\p,
\nonumber\\
&&\smg(1,1),\smg(1,2),\ldots,\smg(3n-2m,1),\smg(3n-2m,2),
(1,1)\p,\smg\p(1),\ldots,(m,1)\p ,\smg(m).
\nonumber
\qqq
Then
\qq
\wgh & = &
(-1)^{|s|}  \eps^{I_{1,0}\ldots I_{2n-m,0} I\p_{1,0}\ldots
I\p_{m,0} }
\prod_{k=1}^{3n-2m} \eps^{I_{\smg(k,1)} I_{\smg(k,2)}}
\prod_{l=1}^m  \eps^{I\p_{l,1} I_{\smg\p(l)} }
\prod_{j=1}^{m-1} \eps^{J\p_{j,1} J\p_{j+1,0} }
\nonumber\\
&&\qquad \times  \Tr_\wa
\prod_{p=1}^{2n-m} \Omega_{I_{p,0}I_{p,1}I_{p,2}I_{p,3}}
\prod_{q=1}^m \Omega_{I\p_{q,0} I\p_{q,1} J\p_{q,0} J\p_{q,1} }.
\label{2.o12}
\qqq
Here $\Tr_\wa$ refers to the uncontracted pair of indices
$J\p_{1,0},J\p_{m,0}$.

\subsection{Perturbative Chern-Simons theory}
\label{s4}

We will want to make a precise comparison of certain sigma-model
invariants to Chern-Simons invariants.  Therefore, we pause to recall
some features of Chern-Simons perturbation theory.

We consider a Chern-Simons theory based on a simple Lie group $G$.
Choose a basis $x_a$ in the Lie algebra of $G$. We denote by
$f_{ab}^{\;\;\;c}$ the structure constants
\qq
[x_a,x_b] = f_{ab}^{\;\;\;c} x_c.
\label{4.1}
\qqq
Let $h_{ab}$ be the Killing form such that the length of long roots
is $\sqrt{2}$. Then the Chern-Simons action on a 3-dimensional
manifold $M$ can be written as
\qq
\Scs = - {1\over 2} \int_M \hdx
\lrbc{
\he h_{ab} A^a_\mu \partial_\nu
A^b_\rho + {1\over 3} \he f_{abc} A^a_\mu A^b_\nu A^c_\rho }.
\label{4.2}
\qqq
The quantum Chern-Simons invariant $\zcs$, $k\in \ZZ$ is expressed as
a path integral
\qq
&
\zcs = {1\over \Vol(\tG) } \int \iA \exp \lrbc{ {i\over \hbcs}
\Scs},
\label{4.3}
\\
& \hbcs = {2\pi \over K}.
\label{4.4}
\qqq
Here $\Vol(\tG)$ is the volume of the group of gauge transformations
and $K= k+\nu$, $\nu$ being the dual Coxeter number of $G$. The
choice of $\hbcs= {2\pi\over K}$ rather than $\hbcs={2\pi\over k}$
is intended to account for a subtle one-loop effect that shifts the
effective value of $k$ seen in perturbation theory relative to what
is seen with different methods of computation.

In the large $K$ approximation, the path integral\rx{4.3} can be
presented as a sum of contributions coming from connected pieces of
the moduli space of flat $G$ connections over $M$. In what follows,
we assume that $M$ is a rational homology sphere, or, in other words,
that
\qq
\dim H_1(M,\IR) =0.
\label{4.5}
\qqq
The trivial connection on such a manifold is an isolated  point in the
moduli space of flat connections. We will use  perturbation theory
 to evaluate the leading large $K$ terms in the
contribution of the trivial connection to the
invariant\rx{4.3}:
\qq
\ztrcs = {1\over \Vol(\tG) } \intr \iA \exp \lrbc{ {i\over \hbcs}
\Scs}.
\label{4.6}
\qqq
Here $\intr$ means that we take the trivial connection contribution
to the path integral in the stationary phase approximation.

A perturbative calculation of the integral\rx{4.6} requires gauge
fixing. We choose the gauge
\qq
\partial^\mu A_\mu = 0.
\label{4.7}
\qqq
We implement the gauge choice by introducing a scalar field $\t^a$
and a pair of fermionic ghost fields $c^a,\bc^a$.
After adding the
gauge fixing terms, the action\rx{4.2} becomes
\qq
\Scsgf & = & \int_M \hdx \lrbc{
- {1\over 2} \he h_{ab} A^a_\mu \partial_\nu A^b_\rho -
{1\over 6} \he f_{abc} A^a_\mu A^b_\nu A^c_\rho
\right.
\label{4.8}\\
&&\left.\hspace*{1in}\qquad\qquad +
h_{ab} \t^a \partial^\mu A^b_\mu +
i h_{ab} \partial^\mu \bc^a \partial_\mu c^b -
if_{abc} \bc^a A^b_\mu\, \partial^\mu c^c }.
\nonumber
\qqq
The propagators of the gauge and ghost fields are
\qq
\langle A^a_\mu(x_1) A^b_\nu (x_2) \rangle & = & -
i\hbcs h^{ab} \Gchi{\mu\nu},
\label{4.9}
\\
\langle c^a(x_1) \bc^b(x_2) \rangle & = &
- \hbcs h^{ab}\Gphi.
\label{4.10}
\qqq

The one-loop correction is known to be
\qq
\ztrocs = {1\over \Vol(G)} \lrbs{
{2\pi\hbcs\over {\ordH}^{1\over 2} } }^{ {1\over 2} \dim G};
\label{4.11}
\qqq
here $\Vol(G)$ is the volume of the gauge group $G$ calculated with
the Killing metric $h_{ab}$. The
contribution of the trivial connection takes the form
\qq
\ztrcs = \ztrocs \exp \lrbs{ \sum_{n=1}^\infty \hbcs^n
S_{n+1}(M) },
\label{4.12}
\qqq
where $S_n(M)$ is the $n$-loop correction calculated according to the
Feynman rules based on propagators\rx{4.9},\rx{4.10} and cubic
interaction vertices
\qq
\vocs & = & -{1\over 6} \he f_{abc} A^a_\mu A^b_\nu A^c_\rho,
\label{4.13}\\
\vtcs & = & - i f_{abc} \bc^a A^b_\mu \, \partial^\mu c^c
\label{4.14}
\qqq
coming from the gauge fixed action\rx{4.8}.

\nsection{Atiyah-Hitchin manifold and Casson-Walker invariant of
rational homology spheres}
\label{sAH}

Our goal here is to make a precise comparison of certain
invariants derived from topological sigma-models with the Casson-Walker
invariant.

The reason that such a relation should exist was explained at the end
of the introduction.  The $N=4$ supersymmetric gauge  theory in
three dimensions, with gauge group $SU(2)$,
 reduces at long distances\cx{SeWi2}
to a supersymmetric sigma
model in which the target space is a certain non-compact \hk
manifold $\ah$.  The Casson-Walker invariant of a three-manifold $M$ is
(apart from a factor of two explained at the end of this section)
the partition function of a topologically twisted form of the $N=4$ gauge
theory on $M$.  It can be computed using any metric.  If one scales
up the metric $h$ on $M$ by $h\to th$ with $t\to \infty$, then one
can replace the gauge theory by the sigma-model -- and the topologically
twisted gauge theory goes over to the topological sigma-model with
target space $\ah$.  The precise $\ah$ that is relevant here is the moduli
space of vacua of the $N=4$ supersymmetric gauge theory.  In
fact, according to\cx{SeWi2}, $\ah$ is the ``reduced moduli space''
of a pair of BPS monopoles on $\R^3$.  It is an asymptotically flat
\hk manifold with fundamental group $\Z_2$; its geometry has been
described in considerable detail in\cx{AH}.

On the other hand, from what we have seen, for {\it any} four-dimensional
\hk manifold $\hx$, the partition function of the topological
sigma-model will be a multiple of the two-loop Chern-Simons invariant.
The Casson-Walker invariant will therefore be a multiple of the two-loop
Chern-Simons invariant, as has been found before in a quite different
way\cx{RoS1},\cx{Ro1}.

We will first write down
 the sigma-model partition function $Z_\hx(M)$ with
a general four-dimensional \hk manifold.
Then we will specialize to the case that
$\hx$ is the manifold $\ah$, and extract the precise statement about the
Casson-Walker invariant.

We will consider the calculation of $Z_\hx(M)$ for four types of
manifolds $M$ for which generally $Z_\hx(M)\neq 0$ - the manifolds
with the first Betti number equal to 0 (\ie rational homology
spheres), 1, 2 and 3.

\subsection{Rational homology spheres}

We start with the sigma-model. There is only one two-loop one-particle
irreducible graph with two trivalent vertices.
We call this graph $\thetx$. Both vertices in the
Feynman diagrams of $\thetx$ can be either $V_1$ or $V_2$.
The corresponding three-manifold integrals are
\qq
\ito(M) & = & \int_M
\epsilon^{\mu_1\mu_2\mu_3} \epsilon^{\nu_1\nu_2\nu_3}
\Gchi{\mu_1\nu_1}\Gchi{\mu_2\nu_2}\Gchi{\mu_3\nu_3} d^3 x_1 d^3 x_2,
\label{5.1}\\
\itt(M) & = & \int_M
h^{\mu_1\mu_2}(x_1) h^{\nu_1\nu_2}(x_2) \Gchi{\mu_1\nu_1}
\label{5.2}\\
&&\qquad\times
\lrbc{ \pp{}{x_1^{\mu_2} } \Gphi }
\lrbc{ \pp{}{x_2^{\nu_2} } \Gphi }
\sqrt{ h(x_1) } d^3 x_1 \sqrt{ h(x_2) } d^3 x_2.
\nonumber
\qqq
The weight function $b_\thetx(\hx)$ is equal to the integral
\qq
b_\thetx(\hx) = {1\over 4\pi^2}
\int_\hx \sqrt{g} d^4 \cvphi^i \,\eps^{I_1J_1} \eps^{I_2J_2}
\eps^{I_3J_3} \eps^{I_4J_4} \Omega_{I_1 I_2 I_3 I_4} \Omega_{J_1 J_2
J_3 J_4} = {1\over 8\pi^2} \Tr \int_\hx R\wedge R.
\label{5.3}
\qqq
Here $R$ is the curvature of $\hx$. The latter integral was
calculated\cx{SSZ}
for the case when $\hx$ is the Atiyah-Hitchin manifold $\ah$:
\qq
b_\thetx(\ah) = {1\over 8\pi^2}
\Tr \int_{\ah} R\wedge R = - 2.
\label{5.4}
\qqq
The authors of\cx{SSZ} integrated over the simply-connected
double cover of $\ah$, so we multiplied their result by $1/2$.

Now let us calculate the combinatorial factors. Consider first the
diagram with 2 vertices $V_1$. Each vertex carries a factor of $-1$
($-1$ comes from $e^{-{1\over \hbar} S}$ in \ex{3.3}, $1\over 6$
comes from \ex{3.3} and $3!$ comes from the fact that $V_1$ has 3
identical legs). The diagram is symmetric under permutations of 2
vertices and 3 propagators, which produces a factor of $1\over 12$.
Rearranging the fermionic fields requires an even number of
permutations, so no sign factor comes from there.
Thus we find
that the diagram with two vertices $V_1$ contributes
\qq
{1\over 12} \ordH \ito(M)
b_\thetx(\hx)
\label{5.5}
\qqq

Consider now the diagram with two vertices $V_2$. Each vertex carries a
factor of $-1$ from $e^{-S}$. The diagram has a symmetry of permuting
the vertices, hence the factor ${1\over 2}$. Rearranging the
fermionic fields produces $-1$. Each propagator $\lrl{ \vphi \vphib}$
carries $-1$.
An extra factor $-1$ comes from arranging the
indices of tensors $\Omega$ into the contraction of \ex{5.3}.
Therefore the diagram with two vertices $V_2$ contributes
\qq
{1\over 2} \ordH \itt(M) b_\thetx(\hx).
\label{5.6}
\qqq
As a result,
\qq
Z_\hx(M) =
{1\over 2} b_\thetx(\hx)
\ordH \lrbc{ {1\over 6} \ito(M) + \itt(M) }.
\label{5.7}
\qqq

A calculation of the two-loop correction $S_2(M)$ to the trivial
connection contribution to the quantum Chern-Simons invariant goes
along the similar lines. First, we consider a diagram with two vertices
$\vocs$. Its contribution is proportional to $\ito(M)$. The
coefficient is composed of the following factors: $-i$ from every
vertex\rx{4.13}, $-i$ from every propagator\rx{4.9},
${1\over 12}$ from
the symmetry of the diagram and
\qq
f^{abc} f_{abc} = 2 \nu \dim G
\label{5.8}
\qqq
from contracting the Lie algebra indices.  Here $\nu$ is the dual
Coxeter number of $G$. Thus the contribution to $S_2(M)$ is
\qq
-{i\over 6} \nu \dim G \;\ito(M).
\label{5.9}
\qqq

Now we turn to the diagram with 2 vertices $\vtcs$. Apart from
$I_2(M)$, we find the following factors: $1$ from every vertex, $-i$
from the propagator\rx{4.9}, $-1$ from permuting the fermions,
$1\over 2$ from the symmetry of the diagram and the factor\rx{5.8}
with extra $-1$ from contracting the Lie algebra indices. Thus the
contribution of the diagram with 2 vertices $\vtcs$ is
\qq
-i \nu \dim G\; \itt(M)
\label{5.10}
\qqq
and, as a result,
\qq
S_2(M) = - i \nu \dim G \lrbc{ {1\over 6} \ito(M) + \itt(M) }.
\label{5.11}
\qqq
Comparing \eex{5.7} and\rx{5.11} we find that
\qq
Z_{\hx}(M) = {i\over 2} b_\thetx(\hx) \ordH {S_2(M) \over \nu \dim G}
\label{5.n12}
\qqq
and, in particular,
\qq
Z_{\ah}(M) = -i\ordH {S_2(M) \over \nu \dim G}.
\label{5.12}
\qqq

It should be noted that the sum of integrals
\qq
{1\over 6} \ito(M) + \itt(M)
\label{5.13}
\qqq
appearing in \eex{5.7} and\rx{5.11}
depends on the metric $h_{\mu\nu}$ of the manifold $M$. This
dependence can be compensated\cx{BNWi} by adding a special counterterm
${1\over 48\pi} \ihfr{\fr}$. Suppose that $M$ is equipped with a local
basis in its tangent bundle. The topological class of the basis
defines a framing $\fr$ of $M$. Let $\omega$ be a Levi-Civita
connection of $M$ relative to this basis. Then
\qq
\ihfr{\fr} = {1\over 4\pi} \int_M \lrbc{
\omega \wedge d\omega + {2\over 3} \omega\wedge\omega\wedge\omega}.
\label{5.14}
\qqq
If the framing is changed by 1 unit then the integral\rx{5.14}
changes by $2\pi$.
Since we expect both $Z_\hx(M)$ and $S_2(M)$ to be topological
invariants, we assume that the counterterm ${1\over 48\pi} \ihfr{\fr}$
is added to the sum\rx{5.13}:
\qq
Z_\hx(M,\fr) & = & -{1\over 2}  b_\thetx(\hx)
\ordH \lrbc{ {1\over 6} \ito(M) + \itt(M) + {1\over
48\pi}\ihfr{\fr} },
\label{5.15}
\\
S_2(M,\fr) & = & i\nu\dim G\lrbc{ {1\over 6}\ito(M) + \itt(M) +
{1\over
48\pi}\ihfr{\fr} }.
\label{5.16}
\qqq

We paid a price for making $Z_\hx(M)$ and $S_2(M)$ topological
invariants: they now depend on the choice of framing. This dependence
is well-known for $S_2(M)$. Fortunately, for any three-manifold $M$ there
is a preferred {\em canonical} framing $\fr_0$. Therefore
$S_2(M,\fr)$ and $Z_\hx(M,\fr)$ can be converted into true topological
invariants of $M$ if they are calculated at $\fr=\fr_0$. A reduction
to the canonical framing is achieved with the help of relations
\qq
Z_\hx(M,\fr) & = & Z_\hx(M) - {1\over 48} b_\thetx(\hx)\ordH \Delta\fr,\qquad
\label{framch}\\
S_2(M,\fr) & = & S_2(M) + {1\over 24} i\nu\dim G \Delta\fr.
\label{framchs}
\qqq
Here $Z_\hx(M)$ and $S_2(M)$ denote the invariants calculated at
the canonical framing, and $\Delta\fr$ is the framing correction:
\qq
\dfr = \fr - \fr_0.
\label{5.21}
\qqq

There is actually an important point that should be made here about the
topological sigma-model.  If this is regarded as a theory in its own
right, it requires some choice of framing, and there is nothing more to say.
But in some cases -- such as the case that the target is the Atiyah-Hitchin
manifold $\ah$ --  the sigma-model can be viewed as the low
energy limit of a model (in this case a supersymmetric gauge theory) with
much softer ultraviolet behavior.  When this is so, as the soft theory
will not generate  a two-loop framing anomaly, the sigma-model will be
endowed
with a natural framing if it is viewed as the low energy limit of the gauge
theory.  It is natural to suspect that this framing is $\fr=\fr_0$,
and we will assume this in writing the following formulas.

Let $\lcw$ be the Casson-Walker invariant of $M$. We will use it also
in a slightly different normalization:
\qq
\lc = \ordH \lcw.
\label{5.18}
\qqq
The normalization of $\lc$ seems to be more in line with
Casson's original definition: apart from the contribution of reducible
flat connections $SU(2)$, it counts the Euler characteristics of the
moduli spaces of irreducible flat $SU(2)$ connections. It is
known\cx{RoS1},\cx{Ro1}
that $S_2(M,\fr)$ calculated at canonical framing is proportional to
$\lcw$:
\qq
S_2(M,\fr_0) = {i\over 2} \nu \dim G\, \lcw(M).
\label{5.17}
\qqq
Therefore \ex{5.12} is compatible with\rx{ahl}
or, more generally, with
\qq
Z_\ah(M,\fr) = {1\over 2} (-1)^{b_1} \lc + {1\over 24} \ordH \dfr,
\label{5.20}
\qqq
where $\dfr$ is the framing correction\rx{5.21}.

The factor of $1/2$ in\rx{ahl} has the following natural interpretation.
In topological $SU(2)$ gauge theory (which comes from the twisting of
the $N=4$ supersymmetric 3-dimensional theory), a gauge
transformation by the constant element $-1$ of the center of $SU(2)$
acts trivially on the space of connections. Therefore, an isolated,
irreducible $SU(2)$ connection is invariant under a group of order
two, consisting of gauge transformations by $\pm 1$, and contributes
$\pm 1/2$, rather than $\pm 1$, to the partition function of the
topological field theory. This factor of $1/2$ is omitted in the
usual definition of the Casson-Walker invariant (although originally
Casson put it in motivated by the fact that the invariant $\lc$
appeared to be always even for integer homology spheres).

\subsection{Manifolds with $b_1(M)=1$}
\label{ssb1}

Let $M$ be a manifold with $b_1(M)=1$. Let $\cb{1}_\mu$ be a one-form
on $M$ which represents the integral cohomology class. The only
minimal Feynman graph contributing to $Z_\hx(M)$ is the loop with two
vertices sitting on it. Both vertices can be either $V_1$ or $V_2$.
The corresponding integrals of Green's functions are
\qq
I_1(M) & = & \int_M
\epsilon^{\mu_1\mu_2\mu_3} \epsilon^{\nu_1\nu_2\nu_3}
\cb{1}_{\mu_1}(x_1) \cb{1}_{\nu_1}(x_2)
\Gchi{\mu_2\nu_2}\Gchi{\mu_3\nu_3}
d^3 x_1 d^3 x_2,
\label{5.b1}\\
I_2(M) & = & \int_M
h^{\mu_1\mu_2}(x_1) h^{\nu_1\nu_2}(x_2)
\cb{1}_{\mu_1}(x_1) \cb{1}_{\nu_1}(x_2)
\label{5.b2}\\
&&\qquad\times
\lrbc{ \pp{}{x_1^{\mu_2} } \Gphi }
\lrbc{ \pp{}{x_2^{\nu_2} } \Gphi }
\sqrt{ h(x_1) } d^3 x_1 \sqrt{ h(x_2) } d^3 x_2.
\nonumber
\qqq
The accompanying weight function is still that of \ex{5.3}. The
combinatorial factor for $I_2$ is the same as for the case of
$b_1(M)=0$. The combinatorial factor for $I_1$ is 3 times bigger,
because the symmetry permutes only two propagators.
As a result,
\qq
Z_\hx(M) = {1\over 2} b_\thetx(\hx) \ordH\p
\lrbc{ {1\over 2} I_1(M) + I_2(M) }.
\label{5.b3}
\qqq

The same combination of integrals as in \ex{5.b3} appears in
Chern-Simons theory on $M$. To be specific, consider a Chern-Simons
theory with the gauge group $SU(2)$. The basis of the Lie algebra
$su(2)$ is formed by Pauli $\sigma$-matrices:
$i\sigma_1, i\sigma_2, i\sigma_3$. The corresponding structure
constants are
\qq
f_{ab}^{\;\;\;c} = -2 \epsilon_{abc}.
\qqq
Here $\epsilon_{abc}$ is an antisymmetric tensor with $\epsilon_{123}=1$.

For an arbitrary real parameter $\tx$, consider a flat connection on
$M$
\qq
A^a_\mu = \delta^a_3\, \tx\, \cb{1}_\mu.
\label{5.b4}
\qqq
The  one-loop contribution of the fields
\qq
A^{1,2}_\mu, c^{1,2}, \bc^{1,2}, t^{1,2}
\label{fields}
\qqq
to the Chern-Simons
partition function at the background\rx{5.b4} is known to be equal to
the inverse Reidemeister torsion
$\lrbs{\treid}^{-1}$. The torsion has the following expansion at
small $\tx$:
\qq
\treid = {1\over 4\tx^2}\,
\ordH\p \; \lrbc{ 1 + \sum_{m=1}^\infty C_n(M) \tx^{2n}
}.
\label{5.b5}
\qqq
Here $C_n(M)$ are some invariants of $M$. The prefactor
$1/4\tx^2$ is
due to the zero modes. The coefficient $-C_1(M)$
is the quadratic contribution to the inverse Reidemeister torsion. It
is calculated by the same diagrams that led to \ex{5.b3}, except that
we have to use either two vertices\rx{4.13} or two vertices\rx{4.14}.
The calculation of combinatorial factors is similar to that for
\ex{5.11} with a few exceptions. $I_1(M)$ acquires  an extra factor of 3
due to the reduction of symmetry between the propagators, there is an
overall extra factor of $i/\hbcs$ because one pairing\rx{4.9} is
substituted by two fields\rx{5.b4}, and we should put
$f_{3ab} f_{3}^{\;ab}=8$ instead of $2\nu\dim G$. As a result,
\qq
-C_1(M) = 4\lrbc{ {1\over 2} I_1(M) + I_2(M) }.
\label{5.b6}
\qqq
Comparing this with \ex{5.b3} we conclude that
\qq
Z_\hx(M) = - {1\over 4}\,b_\thetx(\hx)
\lrbc{ \tx^2 \treido }^{\prime\prime}_{\tx=0}
\label{5.b7}
\qqq
and, in particular,
\qq
Z_{\ah}(M) = {1\over 2}
\lrbc{\tx^2 \treido}^{\prime\prime}_{\tx=0}.
\label{5.b8}
\qqq
This formula is in agreement with \ex{ahl} and the results
of\cx{Lescop}.

If $M=S^2\times S^1$, then
$\treido = {1\over 4\sin^2\lrbc{\tx\over 2} }$ and
\qq
Z_\hx(S^2\times S^1) = - {1\over 24} \,b_\thetx(\hx),\qquad
Z_{\ah}(S^2\times S^1) = {1\over 12}.
\label{5.b9}
\qqq

Another useful example of an application of \ex{5.b8} is a torus
bundle over a circle. This is a 3-manifold constructed by gluing two
$T^2$ boundaries of a 3-manifold $T^2\times [0,1]$ through an
$SL(2,\ZZ)$ twist
\qq
U = \pmatrix{p & r \cr q & s}, \qquad
ps - qr = 1.
\label{tb.1}
\qqq
The Reidemeister torsion of the manifold $T_U$ constructed in this
way is
\qq
\treidxo{T_U} = {e^{-i\tx} \over 4\sin^2\lrbc{\tx \over 2} }
\det \lrbc{ U - e^{i\tx} I}
= { 2\cos\tx - (p+s) \over 4\sin^2\lrbc{\tx \over 2} }.
\label{tb.2}
\qqq
As a result,
\qq
Z_\hx(T_U) = {1\over 24} b_\thetx(\hx)\, (p+s+10).
\label{tb.3}
\qqq

\subsection{Manifolds with $b_1(M)=2$}
\label{btwo}
Let $M$ be a manifold with $b_1(M)=2$. Let $\cb{\ia}_\mu$,
$\ia=1,2$ be the basic integral one-forms. The only minimal
Feynman diagram contributing to $Z_\hx(M)$ consists of two vertices
$V_1$ connected by a single propagator\rx{3.19}. All remaining legs
absorb the zero modes of $\eta^I$ and $\chi^I$. The corresponding
integral is
\qq
I_1(M) & = & \int_M
\epsilon^{\mu_1\mu_2\mu_3} \epsilon^{\nu_1\nu_2\nu_3}
\cb{1}_{\mu_1}(x_1) \cb{1}_{\nu_1}(x_2)
\cb{2}_{\mu_2}(x_1) \cb{2}_{\nu_2}(x_2)
\Gchi{\mu_3\nu_3}
d^3 x_1 d^3 x_2,
\label{5.b14}
\qqq
This can be evaluated in the following way.  Let $c$ be the two-form
$c=\omega^{(1)}\wedge \omega^{(2)}$.  Then as
$\omega^{(1)}\wedge c=\omega^{(2)}\wedge c=0$,
it follows from Poincar\'e duality that the cohomology
class of $c$ is trivial, so that $c=dg$, with $g$ a one-form.  Using
the fact that the propagator $G$ in\rx{5.b14} obeys $dG=\delta$ (where
$\delta$
is a delta function supported on the diagonal in $M\times M$), it follows
that $I_1= \int_Mg\wedge \omega^{(1)}\wedge \omega^{(2)}$, showing
the appearance of the Massey product or Milnor linking number, as promised
above.

The weight function is still $b_\thetx(X)$ and the combinatorial
factor is the one in \ex{5.b3} times $2$ due to the reduction of
symmetry between the propagators.
As a result,
\qq
Z_\hx(M) =  {1\over 2} b_\thetx(X) \ordH\p I_1(M),
\label{5.b15}
\qqq
and, in particular,
\qq
Z_\ah(M) = - \ordH\p I_1(M).
\label{5.b16}
\qqq
The latter result is consistent with the calculations of\cx{Lescop}.
The only difference is that the integral $I_1(M)$ was presented there
in a ``Poincar\'{e} dual'' way: as a self-linking number of the
intersection of two 2-cycles which are dual to the forms
$\cb{\ia}_\mu$.

\subsection{Manifolds with $b_1(M)=3$}

Let $M$ be a manifold with $b_1(M) = 3$. Let $\cb{\ia}_\mu$,
$\ia=1,2,3$ be one-forms giving a basis of the image of $H^1(M,\Z)$ in
$H^1(M,\R)$.   The only minimal
Feynman diagram contributing to $Z_\hx(M)$ consists of two disconnected
vertices
$V_1$, all their legs being attached to the zero modes. The
corresponding integral is the square of the intersection number of the
forms $\cb{\ia}_\mu$:
\qq
I_1 = \lrbc{
\int_M \epsilon^{\mu_1\mu_2\mu_3}
\cb{1}_{\mu_1}(x) \cb{2}_{\mu_2}(x) \cb{3}_{\mu_3}(x)\, d^3 x }^2.
\label{5.b10}
\qqq
This integral is one of the most obvious classical invariants of a
three-manifold
with $b_1(M)=3$.
The weight function is again $b_\thetx(\hx)$. The combinatorial factor
is the same that appeared at $I_1(M)$ in \ex{5.b15} except that we use
the zero-mode expectation value\rx{3.8a5} instead of the remaining
propagator\rx{3.19}. As a result,
\qq
Z_\hx(M) = {1\over 2} \ordH\p b_\thetx(\hx) I_1(M),\qquad
Z_{\ah} = - \ordH\p I_1(M).
\label{5.b12}
\qqq
The latter equation is consistent with \ex{ahl} and with
the calculations of\cx{Lescop}.

Let $M$ be a 3-dimensional torus
$T^3$. Since $|H_1(T^3,\ZZ)|\p=1$ and $I_1(T^3)=1$, we find that
\qq
Z_\hx(T^3) = {1\over 2} b_\thetx(\hx), \qquad
Z_{\ah}(T^3) = - 1.
\label{5.b13}
\qqq

\nsection{Hilbert spaces, operators and gluing properties of the
Casson-Walker invariant}
\label{s6}

In this section, we will begin to analyze the ``physical Hilbert spaces''
obtained by quantizing the topological sigma-model on a Riemann surface
$\Sigma$.
This will give us information about
the gluing properties of the Casson-Walker invariant
in view of its relation to the partition function\rx{3.3} of the
topological sigma-model.  Our analysis will be more complete for
the case that $\Sigma$ has genus zero or one.

Let $M$ be a three-manifold with a boundary $\partial M = \Sigma$;
here $\Sigma$ is a  Riemann surface. The path integral\rx{3.3} for
such an $M$ has to be taken over the fields
$\phi^i,\eta^I,\chi^I_\mu$ which satisfy certain boundary conditions
on $\Sigma$. As a result, $Z_\hx(M)$ becomes a function of the boundary
values. We denote this function as $|M\rangle$. All permissible
functions of the boundary values on $\Sigma$ form a Hilbert
space\footnote{More properly,
in a non-unitary theory such as the twisted
theory considered here, these spaces are vector spaces
with a non-degenerate inner product which is not positive.
It is conventional in physics
to call them Hilbert spaces anyway.}
$\cH_\Sigma^0$. In topological quantum field theories, one can try to
restrict oneself to $Q$-invariant boundary conditions modulo the
action of $Q$. (The present theory has two $Q$'s and one can, for
instance, simply use any linear combination of them.) Equivalently,
one can construct the full Hilbert space and then identify the
cohomology of $Q$ as the physical Hilbert space ${\cal H}_\Sigma$
appropriate to the topological field theory.  The ${\cal H}_\Sigma$
are often finite-dimensional, though the underlying Hilbert space
${\cal H}_\Sigma^0$ of all states, not necessarily $Q$-invariant, is
always infinite-dimensional.

Suppose that the boundaries of two oriented 3-manifolds $M_1$ and
$M_2$ are isomorphic to the same Riemann surface $\Sigma$. We can
glue $M_1$ and $M_2$ together along $\Sigma$, thus obtaining a new
manifold $M$. According to quantum field theory, the partition
function of $M$ can be expressed through a scalar product in the
Hilbert space $\cH_\Sigma$
\qq
Z_\hx(M) = \langle \invor{M_2} | M_1 \rangle =
\langle \invor{M_1} | M_2 \rangle.
\label{scinv}
\qqq
Here $\invor{M}$ denotes a manifold $M$ with the opposite
orientation.

We could twist a boundary of a manifold $M_1$ by an
element $U$ of the  mapping class group of $\Sigma$ prior to gluing.
The mapping
class group (or its central extension) is represented in
$\cH_\Sigma$, and this representation determines the partition
function of the manifold $M^U$ constructed by gluing $M_1$ and $M_2$
after a twist $U$
\qq
Z_\hx(M^U) = \langle \invor{M_2} | U | M_1 \rangle.
\label{surg}
\qqq

The formulas\rx{scinv} and\rx{surg} require some caution due to the
problems related to the sign of the determinant\rx{3.505}. We will
see in the next subsections that the path integral calculation of
the states $|M_1\rangle,\;|M_2\rangle$ most easily determines them
only up to a
sign. A complete definition of these states requires a choice of
orientation on the spaces of the zero modes of the Reidemeister
torsion (with appropriate boundary conditions). For a manifold
with boundary this choice is not canonical. A gluing of $M_1$ and
$M_2$ equipped with orientations on the spaces of zero modes, induces
an orientation on the space\rx{3.8a3}. This induced orientation may
differ from the canonical one. In this case one has to put an extra
negative sign in the \lhs of \eex{scinv} and\rx{surg}.

As we will see shortly, the space $\cH_\Sigma(\hx)$
for $\Sigma=S^2, T^2$ is related to $\bar{\partial}$-cohomology of
certain classes of forms on $\hx$.
If $\hx$ is non-compact, for instance, the Atiyah-Hitchin manifold
$\ah$, then there are several kinds of $\bar\partial$ cohomology
one might consider (\eg
ordinary cohomology, cohomology with compact
support, and ${\bf L}^2$ cohomology)  which give quite different
answers.  In general, the analysis by cutting and summing over
physical states is likely to be quite subtle if $\hx$ is not compact,
roughly because there is a continuum of almost $Q$-invariant states
starting at zero energy.  In the presence of such a continuum,
formal arguments claiming to show a reduction to the $Q$-cohomology
are hazardous at best.
But  if $\hx$ is compact, the spectrum is discrete,
and  one will get a quite straightforward
formalism involving a sum over finitely many physical states.

On the other hand, we know from subection\rw{s3w} that the dependence
on $\hx$ is very simple -- $\hx$ only enters via the values of certain
curvature integrals.  Our approach, therefore, will be to
apply the cutting and pasting formalism for compact $\hx$, and
infer the behavior for general $\hx$  from the values of the curvature
integrals.

For illustrative purposes and application
to the Casson-Walker invariant, we will consider the case
that $\hx$ is four-dimensional.  There is then only one relevant
curvature integral (related to the fact that there is only one
one-particle irreducible trivalent graph with two loops), so
any four-dimensional  compact \hk manifold $\hx$ with
a non-zero value of this one integral can serve as a universal
example.  There is only one candidate: a K3 surface.  Then
if we want to make a statement, for example, about the Casson invariant,
we must take $\hx$ to be the non-compact manifold $\ah$; in going
from K3 to $\ah$ we simply multiply the three-manifold invariant
by a constant, which is the ratio of the appropriate weight
functions. Since
\qq
b_\thetx(\kth) = 48,
\label{bah}
\qqq
we have for any 3-manifold $M$,
\qq
Z_{\kth}(M) =
{b_\thetx(\kth)\over b_\thetx(\ah)} Z_{\ah}(M)
= -{1\over 4} b_\thetx(\kth) \lc(M)
= -12 \lc(M).
\label{bah1}
\qqq

In order to find the structure of $\cH_\Sigma$, we consider a
manifold $M=\IR^1\times \Sigma$ with coordinates $x^0$ and
$(x^1,x^2)$ referring to $\IR^1$ and $\Sigma$. The topological nature
of the sigma-model means that we can try to construct
the physical Hilbert space
 by quantizing the theory in the small $\hbar$ limit. In
doing so, the first step is to find the solutions of the linearized
equations of motion obtained in expanding around a constant map from
$M$ to $\hx$, and then quantize.  Using the  invariance of $M$ under
time translations, one can expand the space of classical solutions
in a basis of modes of definite frequency; by ``zero modes'' and
``non-zero modes'' we mean eigenmodes of zero or non-zero frequency.
In the leading approximation, one gets a Fock space ${\cal F}$
of the non-zero
modes of the various fields tensored with (or rather fibered over)
a more complicated structure built from the zero modes.  As is usual
in such theories, $Q$ is acyclic in ${\cal F}$ away from the ground
state, so we can throw away the non-zero modes and simply quantize
the zero modes.\footnote{An important source of simplification in the
particular models considered here is that, as one varies the constant
map to $\hx$, there are no ``singularities'' at which the separation
between zero and non-zero modes breaks down.  In other examples
such as topological gauge theories related to Donaldson theory,
such singularities are a prime cause of difficulty.}

The $x^0$-independent classical fields for the action\rx{3.5} include
the constant bosonic fields $\phi^i$, constant
fermionic fields $\eta^I, \chi_0^I$ as well as the fields
$\chi^I_\mu(x) = \chi_{\ia}^I\chif{\ia}_\mu$, $\mu=1,2$,
$1\leq \ia \leq \dim H^1(\Sigma)$, here $\chif{\ia}$ are harmonic
one-forms on $\Sigma$, and $\chi_{\ia}^I$ are constant fermionic
coefficients.

The first order structure of the
fermionic part of the action\rx{3.5} indicates that the fermionic
fields satisfy the following anti-commutation relations (we assume
here that $\hbar=1$):
\qq
&\{ \eta^I, \chi_0^J\} = \eps^{IJ},
\label{6.1}\\
&\{ \chi^I_\ia, \chi^J_\ib \} =  - \eps^{IJ}\,
\lrbc{\inter^{-1} }_{\ia\ib}.
\label{6.2}
\qqq
Here $\inter^{\ia\ib}$ is the matrix of
intersection pairing on $\Sigma$:
\qq
\inter^{\ia\ib} = \int_\Sigma \chif{\ia} \wedge \chif{\ib}.
\label{6.3}
\qqq
The formulas\rx{6.1},\rx{6.2} together
with some additional considerations below
lead us to propose the following
structure of the Hilbert space $\cH_{\Sigma_g}$ for a genus $g$
Riemann surface $\Sigma_g$ and a compact \hk manifold $\hx$. We
recall that $\Vbun$ is the natural $Sp(n)$ bundle over $X$.
Let $\Wedge^{\ast} \Vbun$ denote the sum of all exterior powers
of $\Vbun$.
Pick one of the complex structures on $X$.
Then we conjecture that $\cH_{\Sigma_g}$ is the sum of
$\bar{\partial}$-cohomology groups of  $\hx$
with values in the $g$-th tensor power of the bundle
$\Wedge^\ast\Vbun$:
\qq
\cH_{\Sigma_g} = \sum_{q=0}^{{\rm dim}_{\bf C}\,X}H^{q}_{\bar{\partial}}
\lrbc{ \hx, (\Wedge^\ast\Vbun)^{\otimes g} }.
\label{hspace}
\qqq
The basis for this formula will become clear as we analyze the cases $g=0,1$.
To use the $\cH_{\Sigma_g}$
in relation to three-manifolds requires, however, more than the ``additive''
formula just proposed.  One also needs to understand the action of the
mapping class group.  We will analyze this action in detail for $g=1$ but
believe that there may be some subtlety for $g>1$.

The Hilbert space $\hsg$ is $\ZZ_2$-graded. In other words, it splits
into a sum of bosonic and fermionic subspaces, which are $+1$ and $-1$
eigenspaces of the fermionic parity operator $\fp$. The relative grading
of the states inside a given space $\hsg$ is determined by the action
of the fermionic operators $\eta$ and $\chi$. The absolute grading can
be deduced from the consistency between the Feynman diagram and ``Hilbert
space super-trace'' calculations of partition functions of some
3-manifolds. This consistency dictates that the subspace
$H_{\bar{\partial}}^0\lrbc{\hx,(\Lambda^pV)^{\otimes g} }$
has the fermionic parity $(-1)^{1+g}$. Then the fermionic parity of a
subspace
$H_{\bar{\partial}}^q\lrbc{X,\bigotimes_{i=1}^g \Lambda^{l_i} V }$ is
$(-1)^{1+q+g+\sum_{i=1}^g l_i}$. We will see in the next subsections
how the consistency check works in the case of $g=0,1$.

\subsection{The Hilbert space of a 2-sphere and the formula for a
connected sum of 3-manifolds}
\label{sphere}

Let us consider first the case of $\Sigma= S^2$. Since
$\dim H^1(S^2) = 0$, we have to deal only with the fermionic fields
$\eta^I$ and $\chi_0^I$. The commutation relation\rx{6.1} is
represented in a $2^{2n}$-dimensional space. This space contains the
vacuum state $|0\rangle_\eta$ which is annihilated by $\chi^I_0$:
\qq
\chi_0^I |0\rangle_\eta=0,
\label{6.4}
\qqq
and the states produced by the action of operators $\eta^I$:
\qq
\eta^{I_1}\cdots \eta^{I_\ipe} |0\rangle_\eta.
\label{6.5}
\qqq
Including also the bosons, which parametrize the choice of a point
in $\hx$,  the Hilbert space obtained by quantizing the zero modes
is simply the space of sections
of the exterior algebra $\Lambda^*V$ of the
bundle $V\to \hx$.
  Indeed, the state
\qq
|\psi\rangle = \psi_{I_1\ldots I_\ipe} (\phi )
\eta^{I_1}\cdots \eta^{I_\ipe} |0\rangle_\eta
\label{6.6}
\qqq
is interpreted as a section of $\Lambda^\ipe V$.  The inner product
between quantum field theory states becomes in this subspace
the   scalar product between sections of $\Lambda^\ipe V$ and sections
of $\Lambda^{2n-\ipe}V$
defined as
\qq
\langle \psi^{(1)} | \psi^{(2)} \rangle = { 1 \over (2\pi)^{2n} }
\int_\hx \gdphi\,  \eps^{I_1\ldots I_{2n} }
\psi^{(1)}_{I_\ipe\ldots I_1} \psi^{(2)}_{I_{\ipe+1}\ldots I_{2n} }.
\label{6.8}
\qqq
Note that we changed the order of indices in $\psi^{(1)}$.

The action of the operators $Q_A$ can be inferred from the
transformation laws $\delta \phi^i=\gamma^i_{AI}\epsilon^A\eta^I$,
$\delta \eta^I=0$.  This shows that $Q_A$ will act by differentiating
$\phi$ and adding an $\eta$, thus mapping a section of $\Lambda^\ipe V$
to a section of $\Lambda^{\ipe+1}V$.  The action of $Q_A$ is in fact
\qq
Q_A\,
\psi_{I_1\dots I_\ipe}\eta^{I_1}\dots\eta^{I_\ipe}|0\rangle_\eta
= \gamma^i_{I_0A}D_i\,
\psi_{I_1\dots I_\ipe}
\eta^{I_0}\eta^{I_1}\dots\eta^{I_\ipe}|
0\rangle_\eta.
\label{hhh}
\qqq

If out of the \hk structure on $\hx$, we pick a particular
complex structure, then $\Lambda^\ipe V$ can be identified with the
space of $(0,\ipe)$-forms on $\hx$, and one of the $Q$'s becomes the
$\bar\partial$ operator.  If $\hx$ is compact,
the space of physical states
is thus the finite-dimensional space
$${\cal H}_{{ S}^2}=\bigoplus_{\ipe=0}^{2n}H^{0,\ipe}(\hx)$$
(\cf\rx{hspace} for $g=0$)
and the fermionic parity of a subspace $H^{0,\ipe}(\hx)$ is
$(-1)^{1+l}$.

In the case of compact $\hx$, we can immediately determine
the value of the partition function for the three-manifold
$S^2\times S^1$. The same arguments as in\cx{Wi1} indicate that it
is the $\Z_2$-graded (that is, super-) dimension of the physical
Hilbert space, or
\qq
Z_\hx(S^2\times S^1) = \sdim \cH_{S^2} =
\sum_{\ipe=0}^{2n} (-1)^{1+\ipe} \dim H^{0,\ipe}.
\label{6.a1}
\qqq
For instance, if $\hx$ is of dimension four, then according to the
index theorem for the $\bar\partial$ operator, the right hand side is
\qq
Z_\hx(S^2\times S^1)  = - {1\over 192 \pi^2} \int_\hx \Tr (R\wedge R)=
-{1\over 24} b_\thetx(\hx),
\label{6.a2}
\qqq
which is the same as \ex{5.b9}.
For reasons explained above,
\ex{6.a2} is valid also
for $\hx$ non-compact but asymptotically flat.  Notice that, in
general, the \rhs of\rx{6.a2} is not integral; for
instance, if $\hx$ is the Atiyah-Hitchin manifold $\ah$, then using
the value of the curvature integral for $\ah$ from \ex{5.4}, we
get  $Z_\ah(S^2\times S^1)=-1/12$.
This shows that the non-compactness and continuous spectrum are really
essential; there is a real obstruction to representing the system,
for manifolds such as $\ah$, by a finite-dimensional space of
physical states, which would necessarily give an integer
for the partition function on $S^2\times S^1$.

So to proceed, we take $\hx$ to be a K3 manifold.
The $(0,q)$ part of the $\bar\partial$ cohomology is very simple.
$H^{0,0}$ is one-dimensional, represented by a constant function
or equivalently the Fock ground state
\qq
\psi^{(0)} =|0\rangle_\eta.
\label{6.12}
\qqq
Also $H^{0,1}=0$, and $H^{0,2}$ is one-dimensional, represented
by a state that we can think of as
\qq
\psi^{(2)}={\cal O}_\eta(\omx)\psi^{(0)},
\label{6.y12}
\qqq
where ${\cal O}_\eta(\omx)$
is the BRST-invariant operator\rx{2.on1}, and
we choose a $(2,0)$-form $\omx$ to be
\qq
\omx_{I_1 I_2} =
\left\{
\begin{array}{cl}
\displaystyle
 -{2\pi^2\over \Vol(\hx)} \eps_{I_1 I_2} &\mbox{for $\hx={\rm K3}$ (
$\hx$--compact),}\\
\displaystyle
{1\over b_\thetx(\hx)} \eps^{J_1 J_2} \eps^{K_1 K_2} \eps^{L_1 L_2}
\Omega_{I_1 J_1 K_1 L_1} \Omega_{I_2 J_2 K_2 L_2}
&\mbox{for $\hx=\ah$ (
$\hx$--non-compact)}.
\end{array}
\right.
\label{6.x12}
\qqq
According to \ex{6.8}, with this normalization of the form $\omx$,
the scalar product in $\cH_{S^2}$ is
\qq
\langle \psi^{(0)}| \psi^{(0)} \rangle =
\langle \psi^{(2)} | \psi^{(2)} \rangle = 0,
\label{6.13a1}\\
\langle \psi^{(0)} | \psi^{(2)} \rangle =
- \langle \psi^{(2)} | \psi^{(0)} \rangle = 1.
\label{6.13a2}
\qqq

Because of the appearance of the operator ${\cal O}_\eta$,
we really should define a second invariant of a manifold $M$, namely the
path integral with an insertion of  ${\cal O}_\eta$, that
is the path integral
\qq
Z_\hx(M,O_\eta) = \pint \,\,\,\cO_\eta \exp (-S).
\label{6.18}
\qqq
with an insertion of $\cO_\eta$ at some (immaterial) point on
$M$.  The evaluation of this partition function is elementary;
the insertion of $\cO_\eta$ absorbs the $\eta$ zero modes,
so the whole analysis of Section\rw{s3} collapses to the product
of one-loop determinants with no Feynman diagrams at all to correct
them.
Therefore $Z_\hx(M,{\cal O}_\eta)=0$ if $b_1(M)>0$, as there is
then no way to absorb the $\chi$ zero modes, while
if $M$ is a rational homology sphere then,
with the normalization\rx{6.x12},
\qq
Z_\hx(M,{\cal O}_\eta)=|H_1(M,{\bf Z})|.
\label{6.x18}
\qqq
In fact, this formula is valid for any 3-manifold $M$ because of the
definition\rx{3.h1} of $\ordH$.

Now let us use our knowledge of $\cH_{S^2}(\kth)$ in order to determine the
behavior of $Z_{\kth}(M)$ under the operation of connected sum.
We will first derive this behavior  by an argument similar to that
of\cx{Wi1} and then look at path integrals.
Ignore for a moment the definitions\rx{6.12} and\rx{6.y12}. Define
$\psi^{(0)}$ and $\psi^{(2)}$ as elements of a 2-dimensional space
$\cH_{S^2}(\kth)$, which are produced by the sigma-model path
integral taken over a 3-ball $B^3$ without or with an insertion of
$\cO_\eta$:
\qq
| \psi^{(0)} \rangle = | B^3 \rangle,\qquad
| \psi^{(2)} \rangle = | B^3, O_\eta \rangle.
\label{6.b1}
\qqq
If we glue two 3-balls along their boundary $S^2$, then we get a
3-sphere $S^3$.
Therefore the scalar products
$\langle \psi^{(0)} | \psi^{(0)} \rangle$,
$\langle \psi^{(0)} | \psi^{(2)} \rangle$ and
$\langle \psi^{(2)} | \psi^{(2)} \rangle$ are equal to the path
integral over $S^3$ with 0,1 or 2
insertions of $\cO_\eta$. Since
\qq
\lc(S^3) = 0, \qquad |H_1(S^3,\ZZ)| =1,
\label{6.b2}
\qqq
and for any $M$, the path integral with {\it two} insertions of
${\cal O}_\eta$
is zero,
\qq
Z_{\kth}(M,\cO_\eta,\cO_\eta) = 0,
\label{6.b3}
\qqq
we deduce immediately the scalar products\rx{6.13a1},\rx{6.13a2}.
Note that the operator $\cO_\eta$  is antisymmetric, and hence
\qq
Z_{\kth}(S^3, \cO_\eta) = \langle B^3 | B^3,\cO_\eta \rangle =
-\langle B^3, \cO_\eta | B^3 \rangle
\label{6.bb3}
\qqq

Let $B^3\subset M$ be a 3-ball inside a manifold $M$. Since the space
$\cH_{S^2}(\kth)$ is 2-dimensional, the path integral over
$M\setminus B^3$ is a linear combination of the states $\psi^{(0)}$
and $\psi^{(2)}$. To derive the coefficients of this combination, we
observe that gluing $M\setminus B^3$ with $B^3$ restores the manifold
$M$. The scalar products\rx{6.13a1},\rx{6.13a2} imply that
\qq
|M\setminus B^3 \rangle = Z_\kth(M)\, |\psi^{(2)} \rangle +
\ordH \, |\psi^{(0)} \rangle.
\label{6.14}
\qqq
In a similar way one can deduce that the path integral over
$M\setminus B^3$ with one insertion of $\cO_\eta$ is equal to
\qq
|M\setminus B^3, \cO_\eta\rangle = \ordH \, | \psi^{(2)} \rangle.
\label{6.x14}
\qqq
A manifold with the opposite orientation has the same value of
$\ordH$ and an opposite value of $Z_\hx(M)$, so
\qq
|\invor{M}\rangle = -Z_\kth(M)\, |\psi^{(2)} \rangle +
\ordH \, |\psi^{(0)} \rangle, \qquad
|\invor{M}, \cO_\eta\rangle = \ordH \, | \psi^{(2)} \rangle.
\label{6.14inv}
\qqq

Now we can work out the general formula for the partition function
of a connected sum of manifolds.
By definition,
if we glue two manifolds $M_1\setminus B^3$ and $M_2\setminus B^3$
along their common boundary $S^2$
then we get the {\em connected sum} of $M_1$ and $M_2$:
$M=M_1\# M_2$. Since, according to \ex{scinv}, gluing corresponds
to taking the scalar product, we have
\qq
\hspace{-1.5em}
Z_\kth(M_1\# M_2) =
\langle \invor{(M_1\setminus B^3) } | M_2\setminus B^3 \rangle
= \ordHa{M_1} Z_\kth(M_2) + \ordHa{M_2} Z_\kth(M_1).
\label{6.15}
\qqq
This relation
is in line with \ex{bah1} and with the gluing property of
Casson's invariant:
\qq
\lc(M_1\# M_2) = \ordHa{M_1} \lc(M_2) + \ordHa{M_2} \lc(M_1)
\label{6.16}.
\qqq

We can likewise compute the other invariant
$Z_\kth(M_1\# M_2, \cO_\eta)$ of the connected sum
by gluing $M_1\setminus B^3$ with
$M_2\setminus B^3$, the latter containing the operator $\cO_\eta$:
\qq
Z_\kth(M_1\# M_2, \cO_\eta) =
\langle \invor{(M_1\setminus B^3)} | M_2\setminus B^3,
\cO_\eta \rangle = \ordHa{M_1} \ordHa{M_2}.
\label{6.21}
\qqq
This relation is consistent with \ex{6.15} and the gluing property
of the order of the first homology
\qq
\ordHa{M_1\# M_2} = \ordHa{M_1} \ordHa{M_2}.
\label{6.22}
\qqq

Though deduced for $\hx={\rm K3}$, these formulas should hold
if $\hx$ is a more
general four-dimensional \hk manifold such as $\ah$, for reasons
that have been explained.  All that happens if $\hx$ is changed is that
the partition function $Z$ is multiplied by a constant.  Notice,
for instance, that\rx{6.15} is homogeneous in $Z$ and so remains
unchanged if $Z$ is rescaled.

Let us now derive the formulas\rx{6.14},\rx{6.x14} from the
definitions\rx{6.12},\rx{6.y12} by a direct calculation of the path
integral\rx{3.3} over $M\setminus B^3$. We fix the boundary values of
the fields $\phi^i$ and $\eta^I$
\qq
\left.\phi^i(x)\right|_{x\in\partial(M\setminus B^3)} = \cvphi^i,
\qquad
\left.\eta^I(x)\right|_{x\in\partial(M\setminus B^3)} =
\eta^I_0,
\label{6.b4}
\qqq
and decompose the fields $\phi^i(x)$ and $\eta^I(x)$ according to
\ex{3.4} and
\qq
\eta^I(x) = \eta^I_0 + \teta^I(x).
\label{6.b6}
\qqq
The fields $\cvphi^i(x)$ and $\teta^I(x)$ satisfy the zero boundary
conditions
\qq
\left.\vphi^i(x)\right|_{x\in\partial(M\setminus B^3)} = 0,
\qquad
\left.\teta^I(x)\right|_{x\in\partial(M\setminus B^3)} =
0,
\label{6.b7}
\qqq
which exclude the zero modes. The absence of $\teta^I$ zero modes
means that if $b_1(M)=0$, then the path integral has a non-zero
one-loop contribution. Its absolute value is equal to the Reidemeister
torsion of $M\setminus B^3$. However, since the zero-modes of the
Reidemeister torsion of a manifold with a boundary are not paired up
by the Poincar\'{e} duality, the sign cannot be chosen canonically.
To fix a sign, one has to
choose a particular orientation on the space of zero modes. Our
assumption, that the pure one-loop contribution to the path integral
is equal to
$\ordH\,| \psi^{(0)}\rangle$, amounts to such a choice.

To assess the contribution of Feynman diagrams, we substitute the
decomposition\rx{6.b6} into the vertices\rx{3.9},\rx{3.10}. There are
no zero modes to absorb the fields $\teta^I(x)$, so they can be
set equal to zero in the vertices. Thus the contribution to the path
integral over $M\setminus B^3$ comes from the same Feynman diagrams
that contributed to $Z_{\kth}(M)$, except that the fields $\eta^I$ in
their vertices\rx{3.9},\rx{3.10} should be substituted by the
boundary values $\eta_0^I$, and the integral over $d^{4n}\cvphi^i$
should not be taken. As a result, the contribution becomes a function
of the boundary values $\cvphi^i$, $\eta_0^I$. One can easily see
that this function is equal to $Z_{\kth}(M)\psi^{(2)}$ with
$\psi^{(2)}$ defined by \ex{6.x12}, if one uses the second line in
the definition\rx{6.x12} of $\cO_\eta$. Note, however, that both
definitions\rx{6.x12} coincide for $\hx=\kth$. Thus we are ultimately
led to \ex{6.14}. Note also, that since both the pure one-loop and
the Feynman diagram contributions are proportional to the
Reidemeister torsion, their relative sign is fixed and does not
depend on any arbitrary choice.

To derive \ex{6.x14}, one has to substitute the
decomposition\rx{6.b6} into the operator $\cO_\eta$ and set the
fields $\teta^I(x)$ equal to zero. The contribution to the path
integral is purely one-loop. Since it is also proportional to the
Reidemeister torsion, the relative sign of the \rhs in \eex{6.14}
and\rx{6.x14} is fixed.

This completes what we will say in genus zero.

\subsection{The Hilbert space of a 2-torus and the formula for a
Dehn surgery on a knot}
\label{torus}

We now
consider the case  that $\Sigma$ is a two-torus $T^2$.
The space of harmonic one-forms on $T^2$ is
two-dimensional. We choose two basic one-forms $\chif{1},\chif{2}$
such that for a pair of basic one-cycles $C_{1,2}$ of the torus,
\qq
\int_{C_\ib} \chif{\ia} = \delta_{\ia\ib}.
\label{6.23}
\qqq
The corresponding operators $\chi_\ia^I$ satisfy the anti-commutation
relations\rx{6.2}:
\qq
\{ \chi_\ia^I , \chi_\ib^J \} = \eps^{IJ} \epsilon_{\ia\ib}.
\label{6.24}
\qqq
The representation of the relation\rx{6.24} is completely similar to
that of\rx{6.1}.
We let $\chi_2^I$ act by multiplication operators and
we set
\qq
\chi_1^I= \epsilon^{IJ}{\partial\over\partial\chi_2^J}.
\label{weset}
\qqq
The representation space is formed by the vacuum
$|0\rangle_{\chi_2}$ annihilated by $\chi^I_1$:
\qq
\chi_1^I |0\rangle_{\chi_2} = 0,
\label{6.25}
\qqq
together with the states produced by the operators $\chi_2^I$:
\qq
\chi_2^{I_1}\cdots \chi_2^{I_\ipc} |0\rangle_{\chi_2}.
\label{6.26}
\qqq
There is just one crucial difference from the case of $\eta$ zero modes:
to represent the quantum states as functions of $\chi_2$, or functions
of $\chi_1$, requires a choice of a distinguished direction in $T^2$,
so that the mapping class group $SL(2,\Z)$ of $T^2$ will
act quite non-trivially.

The states obtained by quantizing the zero modes now look like
\qq
|\psi\rangle = \psi_{I_1\ldots I_\ipc, J_1\ldots J_\iqc}
(\phi)\,
\chi_2^{I_1}\cdots\chi_2^{I_\ipc}
\eta^{J_1}\cdots\eta^{J_\iqc}\,
 |0\rangle_{\eta\chi}.
\label{6.27}
\qqq
The states, in other words, are sections of
$\wedge^*V\otimes \wedge^*V$.
Upon picking one of the complex structures on $\hx$, $\wedge^*V$ can
be identified with $\Omega^{0,*}(\hx)$, which is also isomorphic,
given the \hk structure on $\hx$, to $\Omega^{*,0}(\hx)$.
One can thus identify $\wedge^*V\otimes \wedge^*V$ as
$\Omega^{*,*}(\hx)$.
The scalar product between the $(\ipc,\iqc)$ and
$(2n-\ipc, 2n-\iqc)$
forms is
defined as
\qq
\langle \psi^{(1)}|\psi^{(2)}\rangle =
{1\over (2\pi)^{2n}}\int_\hx \gdphi\,
\eps^{I_1\ldots I_{2n}} \eps^{\bJ_1\ldots \bJ_{2n}}\,
\psi^{(1)}_{I_\ipc\ldots I_1, \bJ_\iqc \ldots \bJ_1}
\psi^{(2)}_{I_{\ipc+1}\ldots I_{2n}, \bJ_{\iqc+1}\ldots\bJ_{2n}}.
\label{6.29}
\qqq
Again, one of the $Q$'s coincides with the
$\bar\partial$ operator, so if $\hx$ is compact, the space of physical
states is
\qq
\cH_{T^2}(X)=\bigoplus_{\ipc,\iqc=1}^{2n}H^{\ipc,\iqc}(\hx)
\label{spt2}
\qqq
(\cf \ex{hspace}). If $\hx$ is
non-compact, the continuous spectrum starting at zero energy
obstructs a reduction to a description with a finite-dimensional
space of physical states.  We therefore consider only compact $\hx$,
such as $\hx={\rm K3}$, to obtain the surgery formulas.

The mapping class group of $T^2$ is $SL(2,\ZZ)$. A
matrix
\qq
U = \pmatrix{p & r \cr q & s} \in SL(2,\ZZ), \qquad
ps - qr = 1
\label{6.29a1}
\qqq
transforms a pair of bosonic cycles $(C_1, C_2)$ as
\qq
U:\, \pmatrix{C_1\cr C_2} \mapsto
\pmatrix{ p & q \cr r & s } \pmatrix{C_1 \cr C_2}.
\label{6.30}
\qqq
According to \ex{6.23}, the pairs of operators $(\chi_1^I, \chi_2^I)$
are transformed by the same matrix.
For instance,
$\chi_1^I$ maps to $p\chi_1^I+q\chi_2^I$.  We have
represented $\chi^I_2$ by multiplication and $\chi^I_1$ by
$\epsilon^{IJ}\partial/\partial\chi_2^J$.  The state
$|0\rangle_{\chi_2}$ which obeys\rx{6.25} is  mapped by
$U$ to a state
\qq
|0\rangle_{p,q}=p^n\exp\lrbc{
- {q\over p}\, \eps_{IJ}\chi_2^I\chi_2^J }
|0\rangle_{\chi_2}
\label{6.30x}
\qqq
with
\qq
\left(p\epsilon^{IJ}{\partial\over\partial\chi_2^J}+q\chi_2^I\right)
|0\rangle_{p,q}=0.
\label{ugg}
\qqq
As $\chi_2^I$ maps to $r\chi_1^I+s\chi_2^I$, a general
state $f(\chi_2)|0\rangle_{\chi_2}$ is mapped to
\nn
f(r\epsilon^{IJ}(\partial/\partial \chi_2^J)
+s\chi_2^I)|0\rangle_{p,q}.
\label{guggo}
\nnn

As a special case of this, an upper triangular matrix with $q=0$
maps $|0\rangle_{\chi_2}=|0\rangle_{1,0}$ to itself and
maps $f(\chi)|0\rangle_{\chi_2}$, with $f$ of degree $\iq$,
to $(f(\chi)+g(\chi))|0\rangle_{\chi_2}$, where $g$ is of degree
less than $\iq$.
Also, the      matrix
\qq
S=\pmatrix{ 0 & -1 \cr 1 & 0\cr}
\label{tugg}
\qqq
exchanges ``filled states'' with ``empty states,'' mapping
$f(\chi)|0\rangle_{1,0}$ with $f$ homogeneous of degree $\ipc$
to $\tilde f(\chi)|0\rangle_{1,0}$ with $\tilde f$ homogeneous
of degree $2n-\ipc$.  To be somewhat more precise, for a subset of
indices $\sA\in \{1,\ldots,2n\}$, $S$ maps a monomial
$\prod_{I\subset \sA} \chi_2^I$ to a dual monomial
$\pm \prod_{I\not\in \sA}\chi_2^I$.

It is easy to make this explicit for the case that $\hx$ is of dimension
four.  The interesting example is of course $\hx$ a K3 surface.
The Hodge diamond of K3 is well-known:
\qq
\begin{array}{ccccc}
 & &h^{0,0}\\
& h^{1,0} & & h^{0,1}\\
h^{2,0} & & h^{1,1} & & h^{0,2}\\
& h^{2,1} & & h^{1,2}\\
& &h^{2,2}
\end{array}\qquad =
\begin{array}{ccccc}
& & 1\\
& 0 & & 0\\
1 & & 20 & & 1\\
& 0 & & 0\\
& & 1
\end{array}.
\qqq
Here $h^{p,q}$ is the dimension of $H^{p,q}(X)$.

It is obvious from the construction of the space $\cH_{T^2}$ that the
action of $SL(2,\Z) $ does not change the number of $\eta$'s, which
is $\iqc$, so $\bigoplus_\ipc H^{\ipc,\iqc}$ will furnish an
$SL(2,\Z)$ representation for each fixed $\iqc$. From results in the
preceding paragraph, one can see that $SL(2,\Z)$  acts trivially on
$H^{1,1}$, which is twenty-dimensional, but because of the trivial
$SL(2,\Z)$ action  will play only a limited role in what follows.
As for $H^{\ipc,\iqc}$ with $\ipc,\iqc=0,2$, these groups are all
one-dimensional. From the results of the last paragraph, the matrix
$S$ can be seen to map $H^{0,\iqc}$ to $H^{2,\iqc}$ and vice-versa.
Likewise, an upper triangular matrix leaves $H^{0,\iqc}$ invariant
while mapping $H^{2,\iqc}$ to $ H^{2,\iqc}\oplus H^{0,\iqc}$. From
these facts it follows that $H^{0,\iqc}\oplus H^{2,\iqc}$, for each
$\iqc=0,2$,
transforms as the two-dimensional representation of $SL(2,\Z)$.

The knowledge of the structure of the Hilbert space $\cH(T^2)$
allows us to present an alternative way of calculating the partition
function of a torus bundle over a circle $T_U$,
which was considered in the
end of subsection\rw{ssb1}.
{}From the Hilbert space point of view,
the invariant $Z_\kth(T_U)$ can be calculated as a super-trace of the
matrix $U$ represented in $\cH_{T^2}$:
\qq
Z_\kth(T_U) = {\rm STr}_{\cH_{T^2}(\kth)} U 
\label{tb.4}
\qqq
The fermionic parity of a subspace $H^{\ipc,\iqc}(\hx)$ is
$(-1)^{\ipc+\iqc}$,
so the states in $\cH_{T^2}(\kth)$ are bosonic.
Since the Hilbert space $\cH_{T^2}(\kth)$ splits
into 20 trivial and 2 fundamental representations of $SL(2,\ZZ)$,
we have  according to \ex{tb.4},
\qq
Z_\kth(T_U)= 2(p+s) + 20,
\label{tb.4x1}
\qqq
which is in full agreement with \eex{tb.3} and\rx{bah}. It is attractive
that the number $20={\rm dim}\,H^{1,1}(X)$ appears here.  In particular,
when $U=I$, the manifold $T_U$ is isomorphic to $T^2\times S^1$, and
\ex{tb.4x1} tells us that
\qq
Z_\kth(T^2\times S^1) = 24.
\label{tb.4x2}
\qqq
This is in agreement with  \eex{5.b13} and\rx{bah}.

Let us choose as follows  vectors $\psi^{(\ipc,\iqc)}$, $\ipc,\iqc=0,2$
in the spaces $H^{\ipc,\iqc}$. Choose $\psi^{(0,0)}$
and $\psi^{(0,2)}$  as in the previous subsection:
\qq
\psi^{(0,0)} = 1,\qquad \psi^{(0,2)}_{\bI_1 \bI_2} =
g_{\bI_1 J_1} g_{\bI_2 J_2} \eps^{J_1 K_1} \eps^{J_2 K_2}
\omx_{K_1 K_2};
\label{6.t1}
\qqq
here the $(2,0)$-form $\omx_{IJ}$ is defined by \ex{6.x12}. Define the forms
$\psi^{(2,0)}$ and $\psi^{(2,2)}$ as
\qq
\psi^{(2,0)}_{I_1 I_2} = - {1\over 2} \eps_{I_1 I_2}  \psi^{(0,0)},
\qquad
\psi^{(2,2)}_{I_1 I_2, \bJ_1 \bJ_2} = - {1\over 2} \eps_{I_1 I_2}
\psi^{(0,2)}_{\bJ_1 \bJ_2}.
\label{6.t2}
\qqq
The form $\psi^{(2,2)}$ can be expressed equivalently as
\qq
\psi^{(2,2)}_{I_1 I_2, \bJ_1 \bJ_2} =
{1\over b_\thetx(K3)} g_{\bJ_1 N_1} g_{\bJ_2 N_2}
\eps^{N_1 K_1} \eps^{N_2 K_2} \eps^{L_1 L_2} \eps^{M_1 M_2}
\Omega_{I_1 K_1 L_1 M_1} \Omega_{I_2 K_2 L_2 M_2}.
\label{6.t3}
\qqq
The scalar products of the states\rx{6.t1} and\rx{6.t2} are
calculated with the help of \ex{6.29}:
\qq
&
\langle \psi^{(0,0)} | \psi^{(2,2)} \rangle =
\langle \psi^{(2,2)} | \psi^{(0,0)} \rangle = 1,
\nonumber\\
&
\langle \psi^{(2,0)} | \psi^{(0,2)} \rangle =
\langle \psi^{(0,2)} | \psi^{(2,0)} \rangle = -1.
\label{6.33}
\qqq
The transformation\rx{6.29a1} acts on these basis vectors as
\qq
U:\; \pmatrix{ \psi^{(0,\ast)} \cr \psi^{(2,\ast)} }
\mapsto \pmatrix{ p & q \cr r & s }
\pmatrix{ \psi^{(0,\ast)} \cr \psi^{(2,\ast)} },
\label{6.34}
\qqq

Now we can use Feynman diagrams in order to calculate the states
produced on the torus boundary of a 3-manifold. Let $\cK$ be a knot
in a 3-manifold $M$. For the sake of simplicity, we assume that $M$
is a rational homology sphere and that the homology class of $\cK$
is trivial. We want to determine a state $|M\setminus \cK\rangle$
produced by the path integral over $M$ minus a tubular neighborhood
of $\cK$.

First of all, we have to set the boundary conditions on
$T^2 = \partial(M\setminus \cK)$. We make the following choice of basic
cycles on $T^2$: $C_1$ is the meridian of the knot complement, that
is, the cycle which is contractible through $M\setminus \cK$. $C_2$
is the parallel of the knot complement, that is, the cycle which is
contractible through the tubular neighborhood of $\cK$. $C_1$ and $C_2$
have a unit intersection number. We impose the following boundary
conditions on the fields:
\qq
&\left.\phi^i(x)\right|_{\btor}=\cvphi^i,\qquad
\left.\eta^I(x)\right|_{\btor}=\eta^I_0,\qquad
\nonumber\\
&\left.\chi_\mu^I(x)\right|_{\btor}=
\chivar^I\chif{2}_\mu(x),\qquad \mu=1,2,
\label{6.ty1}
\qqq
Here $\chif{2}$ is one of the forms\rx{6.23} and
$\mu = 1,2$ means that we fix only the components of $\chi^I_\mu(x)$
which are (co-)tangent to $T^2$.

We off-set the boundary conditions by splitting the fields according to
\eex{3.4},\rx{6.b6} and
\qq
\chi^I_\mu(x) = \chivar^I \chif{2}_\mu(x) + \tchi^I_\mu(x).
\label{6.ty2}
\qqq
Here we use the same notation $\chif{2}_\mu$ for a closed one-form
which is an extension of $\chif{2}_\mu$ from $T^2$
to $M\setminus \cK$.
The fields $\vphi^i(x),\teta^I(x), \tchi^I_\mu(x)$ satisfy  boundary
conditions
\qq
\left.\vphi^i(x)\right|_{\btor}=0,\qquad
\left.\teta^I(x)\right|_{\btor}=0,\qquad
\left.\tchi_\mu^I(x)\right|_{\btor}=
0,\qquad \mu=1,2,
\label{6.ty3}
\qqq
which exclude the zero modes. Therefore the one-loop contribution to the
path integral over $\mmk$ is equal to $\ordH$ and the pure one-loop
contribution to the state $|\mmk\rangle$ is
$\ordH\,|\psi^{(0,0)}\rangle$.

Apart from the purely one-loop contribution to $|\mmk\rangle$,
there are two contributions which come from Feynman diagrams. The first
contribution comes from the $\theta$-graph. The fields $\eta^I(x)$ in
its vertices are substituted by the boundary values $\eta_0^I$ coming from
\ex{6.b6}, while the fields $\chi^I_\mu(x)$ are substituted by the fields
$\tchi^I_\mu(x)$ of \ex{6.ty2}. A contribution of this graph was considered
in the previous subsection, it is equal to
$Z_{\kth}(M) \ketpsi{0,2}$.

The second Feynman diagram contribution comes from the same graph as in
subsection\rw{btwo}, except that the fields $\eta^I(x)$ in both of
its vertices are substituted by  $\eta_0^I$, one field $\chi^I_\mu(x)$
in each vertex is substituted by $\chivar^I \chif{2}_\mu(x)$ and
the remaining fields $\chi^I_\mu(x)$ are substituted by $\tchi^I_\mu(x)$.
The contribution of this Feynman diagram can be computed with the help
of the same trick that was used in subsection\rw{btwo}.
Namely, this diagram appears in the calculation of the contribution of the
flat connection
\qq
A^a_\mu = \delta^a_3 \, \tx \, \chif{2}_\mu(x)
\label{6.ty4}
\qqq
(\cf \ex{5.b4}) into the Chern-Simons partition function of the knot
complement $\mmk$. As explained in\cx{EMSS}, this partition function
produces the Jones polynomial of the knot $\cK$ with the
$(\tx K)$-dimensional
representation of $SU(2)$ assigned to it. The graph that we need represents
a part of the one-loop contribution of the connection\rx{6.ty4} that comes
from the fields\rx{fields} and is proportional to $\tx^2$ after the
contribution of the zero modes is factored out. The full one-loop
contribution
of the fields\rx{fields} is equal to the inverse Reidemeister torsion of
$\mmk$:
$\lrbs{\treidmk}^{-1}$ (see,\eg\cx{Ro1} and references therein).
The expansion of the Reidemeister torsion at small $\tx$ is
\qq
\treidmk= {\ordH \over 2i\tx}
\lrbc{1 + \sum_{m=1}^\infty C_n(\mmk)\, \tx^{2n} }.
\label{6.ty5}
\qqq
The prefactor $1/(2i\tx)$ is due to zero modes, so the quadratic
contribution to the inverse Reidemeister torsion is equal to
$-C_1(\mmk)$. After the same algebra as in subsection\rw{btwo}, we conclude
that the remaining contribution to the state $|\mmk\rangle$ is
\qq
- {1\over 4} b_\thetx(\kth)
\lrbc{i\tx\treidomk}^{\prime\prime}_{\tx=0}
\ketpsi{2,2}
\label{6.ty6}
\qqq
(the state $\ketpsi{2,2}$ is produces in its form\rx{6.t3}).

The Reidemeister torsion of the knot complement is related to the
Alexander polynomial of the knot:
\qq
\treidta = {\ordH\over \ta^{1/2} - \ta^{-1/2} }\, \Al.
\label{6.ty7}
\qqq
Therefore the state\rx{6.ty6} can be expressed in terms of the second
derivative
$\Alder = \lrbc{ \Al }^{\prime\prime}_{\ta=1}$ as
\qq
{1\over 4} b_\thetx(\kth) \ordH  \Aldersh \ketpsi{2,2}.
\label{6.ty8}
\qqq
After adding up all three contributions, we find that the state
created by a complement of a knot $\cK$ inside a rational homology
sphere $M$ is
\qq
\lefteqn{
|\mmk\rangle  = \ordH\, \ketpsi{0,0} + Z_\kth(M) \ketpsi{0,2}
}
\label{6.ty9}\\
&&\qquad  \hspace*{2.2in}
+ {1\over 4} b_\thetx(\kth) \ordH \Aldersh \ketpsi{2,2}.
\nonumber
\qqq

Let us insert an operator $\cO_\eta$ into the knot complement $\mmk$.
The only contribution to the state $|\mmk,\cO_\eta\rangle$ is purely
one-loop, and both fields $\eta^I(x)$ in $\cO_\eta$ should be
substituted by $\eta_0^I$. Therefore
\qq
|\mmk, \cO_\eta\rangle = \ordH \, \ketpsi{0,2}.
\label{6.ty10}
\qqq

If we reverse the orientation of $M$, then $Z_\kth(M)$ changes its
sign. Thus
\qq
\lefteqn{
|\invor{(\mmk)}\rangle   =
\ordH\, \ketpsi{0,0} - Z_\kth(M) \ketpsi{0,2}
}\hspace*{1.8in}
\label{6.ty11}\\
\qquad  \hspace*{0.7in}
&\hspace*{0.5in}
+ {1\over 4} b_\thetx(\kth) \ordH \Aldersh \ketpsi{2,2}.
\nonumber\\
&\hspace*{-1.7in}
|\invor{(\mmk)}, \cO_\eta\rangle  =  \ordH\, \ketpsi{0,2}.
\label{6.ty12}
\qqq

Let us use the formulas\rx{6.ty9}--\rx{6.ty12} together with the
scalar product\rx{6.33} and $SL(2,\ZZ)$ representation\rx{6.34} in
order to derive the surgery properties of the partition function
$Z_\kth(M)$. First of all, we apply \ex{6.ty9} to the complement of
an unknot in $S^3$. This complement is isomorphic to a solid torus
$D^2\times S^1$, so
\qq
|D^2\times S^1\rangle = \ketpsi{0,0} - {1\over 48}\, b_\thetx(\kth)\,
\ketpsi{2,2}.
\label{6.ty13}
\qqq
If we glue two solid tori together in an obvious way, then we obtain
$S^2\times S^1$. Since $\invor{(D^2\times S^1)}$ is isomorphic to
$D^2\times S^1$, then according to \ex{scinv},
\qq
Z_\kth(S^2\times S^1) = \langle D^2\times S^1 | D^2\times S^1 \rangle
= - {1\over 24}\, b_\thetx(\kth).
\label{6.ty14}
\qqq
This result is consistent with \ex{5.b9}.

Now let us perform a rational $p/q$ surgery on a homologically
trivial knot $\cK$ in a rational homology sphere $M$. We take a solid
torus $D^2\times S^1$, twist its boundary by a product of matrices
$S U$ defined by \eex{6.29a1} and\rx{tugg}, and then glue it to
the boundary of $\mmk$, thus constructing a new closed manifold
$M\p$. The matrix $S$ is needed to reconcile the definitions of
a meridian and a parallel for the boundaries of $\mmk$ and
$D^2\times S^1$. The partition function of the new manifold $M\p$ is
a matrix element of $SU$:
\qq
Z_\kth(M\p,\fr\p) & = & \langle \invor{(\mmk)} | S U | D^2\times
S^1 \rangle
\label{6.ty15}\\
& = & p Z_\kth(M) - {1\over 4} b_\thetx(\kth) \ordH
\lrbs{ q\Aldersh + {1\over 12} r}.
\nonumber
\qqq

The presence of the framing symbol $\fr\p$ in the \lhs of
\ex{6.ty15} reflects the fact that if the original manifold $M$ has a
canonical framing, then the framing $\fr\p$ of the new manifold $M\p$
is not necessarily canonical. In fact, the framing
correction\rx{5.21} for the $U$ surgery is
\qq
\Delta \fr\p = -12 s(p,q) + {p+s\over q} - 3 \prosign(pq).
\label{6.39}
\qqq
Here $s(p,q)$ is the Dedekind sum. Therefore in order to test the
relation\rx{bah1}, in accordance with \ex{framch}, we have to compare
\ex{6.ty15} with the following modification of Walker's surgery
formula:
\qq
\lefteqn{
\lc(M\p) + {1\over 12} \ordHa{M\p} \Delta\fr\p} \hspace*{1in}
\nonumber\\
& = &
\prosign(p) \left\{p\lc(M) +
 \ordH\lrbs{q\Aldersh + {1\over 12} r} \right\}.
\label{6.40}
\qqq
Comparing \eex{bah1},\rx{6.ty15} and\rx{6.40}, we see that they are
compatible except for the sign factor $\prosign(p)$ which is missing
in \ex{6.ty15}. This means that if we do a $U$ surgery with $p < 0$,
then there is an extra $-1$ appearing in \ex{5.20}.

We suggest the following explanation for the ``anomalous'' sign factor
$\prosign(p)$. The relation\rx{5.20} between the partition function
$Z_{\ah}(M,\fr)$ and the Casson invariant $\lc$, as derived in
Section\rw{sAH}, depends on the choice of the sign for the vacuum
expectation values\rx{3.8a4} and\rx{3.8a5}. This choice is related to
a choice of orientations on the spaces\rx{3.s1}. Although these
spaces have canonical orientations, a surgery on a manifold $M$ with
a canonical orientation dictates an orientation for the spaces on
the new manifold $M\p$. If $p<0$ (or if $p=0$ and
$q>0$), then the surgery-induced orientation differs from the
canonical one, and the partition function $Z_{\ah}(M,\fr)$ picks up
an extra negative sign. This situation is reminiscent of the framing
anomaly of the Chern-Simons gauge theory.

That the sign factor of the origin just described is precisely
${\rm sign}(p)$
should be demonstrated directly, but we will instead just test this claim by
 a surgery calculation of the more elementary invariant $\ordH$.
Applying \ex{6.ty10} to a solid torus $D^2\times S^1$, we find that
\qq
|D^2\times S^1, \cO_\eta\rangle = \ketpsi{0,2}.
\label{6.ty17}
\qqq
Then a combination of \eex{surg} and\rx{6.ty12} leads to the formula
\qq
Z_\kth(M\p, \cO_\eta) =
\langle \invor{(\mmk)} | S U | D^2\times S^1, \cO_\eta \rangle
= p\ordH.
\label{6.42}
\qqq
The surgery formula for $\ordHa{M\p}$ is
\qq
\ordHa{M\p} = \prosign(p)\, p\, \ordH,
\label{6.43}
\qqq
Comparing \eex{6.x18},\rx{6.42} and\rx{6.43} we see that the same
factor $\prosign(p)$ is missing also in \ex{6.42}. This happens for
the same reason as in \ex{6.ty15}.

\bigskip
\noindent
{\bf Acknowledgements.}
We would like to acknowledge helpful discussions with
S.~Axelrod and D.~Freed. The work of
E.W. was supported in part by the NSF Grant PHY95-13835. The work of
L.R. was supported in part by the NSF Grant DMS 9304580.

\nappendix{}
\def\tGamma{ \tilde{\Gamma} }
\def\tR{ \tilde{R} }
\def\tnabla{ \tilde{\nabla} }
\def\tT{ \tilde{T} }
\def\bp{ \bar{\partial} }

After this paper was submitted, M.~Kontsevich\cx{Kol} and
M.~Kapranov\cx{Ka} showed that the
definition of the weight functions $\bg(\hx)$ requires less
structure on the manifold $\hx$ than we have assumed so far. Namely,
$\hx$ does not necessarily have to be \hk. It is enough
for $\hx$ to be complex and to have a holomorphic symplectic
structure. The purpose of this appendix is to explain how this
generalization can be understood in the framework of the topological
sigma-model.

We will construct a  sigma-model whose target is a complex manifold $\hx$ which
reduces in the \hk case to the construction given in Section 2.
Given a choice of local complex coordinates on $\hx$, a map from
a three-manifold $M$ to $\hx$ is described by
 bosonic fields $\phi^I(x^\mu)$ and
$\phib^{\bI}(x^\mu)$ describing the map from the 3-manifold $M$ to
$\hx$ in local coordinates. The model will also have two fermionic fields
$\chi^I_\mu(x^\mu)$
and $\eta^{\bI}(x^\mu)$ which are respectively a one-form and zero-form
which take values in the fibers of the
pull-back of the holomorphic and anti-holomorphic tangent bundles of
$\hx$ respectively.  (To compare with the construction given in Section 2 for
the \hk case, the reader should note that on a \hk manifold
there is a natural isomorphism between the holomorphic and anti-holomorphic
tangent bundles of $\hx$.)

We will construct a model with a single fermionic symmetry that we will
call  $\bQ$.  Its action on
the fields will be
\qq
\begin{array}{lll}
\delta \phi^I = 0, &\qquad &
\delta \phib^{\bI} = \eta^{\bI}, \\
\delta \eta^I = 0, &\qquad &
\delta \chi_\mu^I = - \partial_\mu \phi^I
\end{array}
\label{A.2}
\qqq

We will introduce $\bQ$-invariant lagrangians $L_1$ and $L_2$ that parallel
those of Section 2.
The construction of $L_2$ requries some
extra structure on $\hx$. Let $\Gamma^I_{JK}$ be a symmetric
connection in the holomorphic tangent bundle of $\hx$:
$\Gamma^I_{JK} = \Gamma^I_{KJ}$.
The (1,1)-part of the curvature associated with
$\Gamma^I_{JK}$ represents the so-called Atiyah class of $\hx$
\qq
R^I_{\;JK\bL} = {\partial \Gamma^I_{JK} \over \partial \phib^{\bL} }.
\qqq
We also assume that $\hx$ has a holomorphic symplectic structure
$\eps_{IJ}$.
The (2,0)-form $\eps_{IJ}$ does not have to be covariantly constant
with respect to the connection $\Gamma^I_{JK}$.
We will only use the fact that $\eps_{IJ}$ is
non-degenerate and
closed
\qq
\pp{ \eps_{IJ} }{ \phib^{\bK} }=0, \qquad
\pp{\eps_{IJ}}{\phi^K} + \pp{ \eps_{KI}}{\phi^J} +
\pp{ \eps_{JK} }{\phi^I}=0.
\label{A.1}
\qqq

A $\bQ$-invariant lagrangian $L_2$ is a slight modification of the
lagrangian\rx{2.3}
\qq
L_2 & = & {1\over 2} {1\over \sqrt{h} } \eps^{\mu\nu\rho}
\lrbc{\eps_{IJ}
\chi_\mu^I \nabla_\nu
\chi_\rho^J -
{1\over 3} \eps_{IJ} R^J_{\;KL\bM}
\chi_\mu^I \chi_\nu^K \chi_\rho^L \eta^{\bM}
+ {1\over 3} (\nabla_L \eps_{IK} ) (\partial_\mu \phi^I) \chi_\nu^K
\chi_\rho^L }.
\label{A.3}
\qqq
Here $\nabla_\mu$ is a covariant derivative with respect to the
pull-back of the connection $\Gamma^I_{JK}$
\qq
\nabla_\mu \chi_\nu^I = \partial_\mu \chi_\nu^I +
(\partial_\mu \phi^J) \Gamma^I_{JK} \chi_\nu^K.
\label{A.03}
\qqq
The BRST-class of $L_2$ is independent of the choice of
connection $\Gamma^I_{JK}$. Indeed, if we change a connection
$\Gamma^I_{JK}$ by a
tensor $A^I_{JK}$
\qq
\Gamma^I_{JK} \rightarrow \Gamma^I_{JK} + A^I_{JK},
\label{A.4}
\qqq
then $L_2$ is changed by a BRST-exact term
\qq
L_2 \rightarrow L_2 + \delta \lrbc{
{1\over 3} \eps_{IJ} A^J_{KL} \chi^I_\mu \chi^K_\nu \chi^L_\rho }.
\label{A.5}
\qqq

A construction of the
$\bQ$-exact lagrangian $L_1$
requires a
choice of an hermitian metric $g_{I\bJ}$ on $\hx$. We use the
notations
\qq
\tGamma^{\bI}_{\bJ\bK} = {1\over 2}\, g^{\bI L}
\lrbc{ \pp{g_{L\bJ}}{\phib^{\bK}}
+ \pp{g_{L\bK}}{\phib^{\bJ}}
},\qquad
\tT^{\bI}_{\bJ\bK} = {1\over 2}\, g^{\bI L}
\lrbc{ \pp{g_{L\bJ}}{\phib^{\bK}}
- \pp{g_{L\bK}}{\phib^{\bJ}}
},
\label{A.9}
\qqq
here $\tGamma^{\bI}_{\bJ\bK}$ is a symmetric connection and
$\tT^{\bI}_{\bJ\bK}$ is a torsion associated with $g_{I\bJ}$.
Then $L_1$ can be
written as a BRST commutator
\qq
L_1  =  \bQ\lrbc{ g_{I\bJ} \chi_\mu^I (\partial_\mu \phib^{\bJ} ) }
 =  g_{I\bJ}\, \partial_\mu \phi^I \partial_\mu \phib^{\bJ} +
g_{I\bJ} \chi^I_\mu \tnabla_\mu \eta^{\bJ},
\label{A.10}
\qqq
here $\tnabla_\mu$ is a covariant derivative with respect to the
conenction $\tGamma^{\bI}_{\bJ\bK} + \tT^{\bI}_{\bJ\bK}$
\qq
\tnabla_\mu \eta^{\bI} =
\partial_\mu \eta^{\bI} + (\partial_\mu\phib^{\bJ})
(\tGamma^{\bI}_{\bJ\bK} + \tT^{\bI}_{\bJ\bK})
\eta^{\bK}.
\label{A.11}
\qqq

If $g_{I\bJ}$ is a \k metric, then $\tT^{\bI}_{\bJ\bK}=0$ and the
perturbative calculations in the sigma-model are similar to
those of Section\rw{s3}. In particular, the relevant interaction
vertices contained in the largrangians\rx{A.3} and\rx{A.10} are
\qq
V_1 & = &
-{1\over 6} {1\over \sqrt{h} } \eps^{\mu\nu\rho}
\eps_{IJ} R^J_{\;KL\bM}
\chi_\mu^I \chi_\nu^K \chi_\rho^L \eta^{\bM}
\label{A.12}\\
V_2 & = & g_{I\bJ} \tR^{\bJ}_{\;\bK\bL M} \chi^I_\mu \eta^{\bL}
(\partial_\mu \vphib^{\bK}) \vphi^M,
\label{A.13}
\qqq
here $\tR^{\bI}_{\;\bJ\bK L}$ is the curvature tensor associated with
the connection $\tGamma^{\bI}_{\bJ\bK}$
\qq
\tR^{\bI}_{\;\bJ\bK L} = \pp{\tGamma^{\bI}_{\bJ\bK}}{\phi^L}.
\label{A.14}
\qqq

The formula for the weight function $\bg(\hx)$ is derived from the
vertex\rx{A.13} similarly to how it was done in subsection\rw{s3w}.
The holomorphic indices of the curvature tensors $R^I_{\;JK\bL}$
assigned to the vertices of a closed $2n$-vertex graph
$\Gd\in\Gs_{n,3}$, are contracted with the (inverse) symplectic form
$\eps^{IJ}$. After anti-symmetrizing over the anti-holomorphic
indices, we obtain a $\bp$-closed $(0,2n)$-form on $X$. In
other words, we get a map
\qq
\Gs_{n,3} \rightarrow H^{2n}_{\bp}(X).
\label{A.15}
\qqq
In order to get a weight $\bg(\hx)$ we wedge-multiply the
$(0,2n)$-form, which is the image of a graph $\Gd$, by the
$(2n,0)$-form $\epsn^{I_1\ldots I_{2n} }$ and integrate the resulting
$(2n,2n)$-form over $\hx$.

The weight function $\bg(\hx)$ satisfies the IHX relation, because
the $(0,2n)$-form which corresponds to the image of an IHX
combination of graphs, is $\bp$-exact and thus belongs to the kernel
of the map\rx{A.15}. This follows from one of the Bianchi identities
satisfied by the curvature tensor $R^I_{\;JK\bL}$ which
shows that the Atiyah class is a $\bp$-closed form
\qq
\pp{}{\phib^{\bM}} R^I_{\;JK\bL} =
\pp{}{\phib^{\bL}} R^I_{\;JK\bM}.
\label{A.6}
\qqq
This identity implies that
\qq
\bar{\partial} \lrbc{ \nabla_M R^I_{\;JK\bL} } & \equiv &
\pp{}{\phib^{\bN}} \lrbc{ \nabla_M R^I_{\;JK\bL} } -
(\bN \leftrightarrow \bL)
\nonumber\\
&=&
\lrbs{ \pp{}{\phib^{\bN}},\nabla_{M} }R^I_{\;JK\bL}
+ \nabla_M \lrbc{ \pp{}{\phib^{\bN}} R^I_{\;JK\bL} } -
(\bN \leftrightarrow \bL)
\nonumber\\
&=&
R^I_{\;PM\bN} R^P_{\;JK\bL} - R^P_{\;JM\bN} R^I_{\;PK\bL}
- R^P_{\;KM\bN} R^I_{\;JP\bL} - (\bN \leftrightarrow \bL)
\label{A.7}\\
&=&
R^I_{\;PM\bN} R^P_{\;JK\bL} + R^I_{\;PK\bN} R^P_{\;JM\bL}
+ R^I_{\;JP\bN} R^P_{\;KM\bL} - (\bN \leftrightarrow \bL),
\nonumber
\qqq
that is, the \rhs of \ex{A.7} is $\bar{\partial}$-exact.


\begin{thebibliography}{99}
%
\bimn{Wi1}{E.~Witten}{Quantum field theory and the Jones
polynomial}{Commun. Math. Phys.}{121}{1989}{351-399}
%
\bibitem{AxSi} S. Axelrod, I. Singer, {\em Chern-Simons perturbation
theory}, Proceedings of XXth Conference on Differential Geometric
Methods in Physics (New York, 1991) (S.~Catto and A.~Rocha, eds)
World Scientific, 1992, 3-45.
%
\bimn{AxSi1}{S.~Axelrod, I.~Singer}{Chern-Simons perturbation theory
II} {Journ. Diff. Geom.}{39}{1994}{173-213}
%
\bimn{BN1}{D.~Bar-Natan}{Perturbative Chern-Simons theory}{Journ. of
Knot Theory and its Ramifications}{4-4}{1995}{503-548}
%
\bibitem{Ko}M.~Kontsevich, {\em Feynman diagrams and low-dimensional
topology}, in Proceedings of the first European Congress of
Mathematics, vol.~2, Progress in Math. {\bf 120}, Birkh\"{a}user,
Boston, 1994, 97-121.
%
\bibitem{AxSi2}S.~Axelrod, {\em Overview and warmup example for
perturbation theory with instantons}, preprint hep-th/9511196, to be
published in ``Proceedings on Geometry and Physics''.
%
\bimn{FG}{D.~Freed, F.~Gompf}{Computer calculation of Witten's
3-manifold invariant}{Commun. Math. Phys.}{141}{1991}{79-117}
%
\bimn{RoS1}{L.~Rozansky}{A large $k$ asymptotics of Witten's
invariant of Seifert manifolds}{Commun. Math.
Phys.}{35}{1994}{5219-5246}
%
\bimn{Ro1}{L.~Rozansky}{A contribution of the trivial connection to
the Jones polynomial and Witten's invariant of 3d manifolds}{Commun.
Math. Phys.}{175}{1996}{275-296}
%
\bimn{Oh1}{T.~Ohtsuki}{Finite type invariants of integral homology
3-spheres}{J. Knot Theory and its Rami.}{5}{1996}{101-115}
%
\bimn{BN2}{D.~Bar-Natan}{On the Vassiliev knot
invariants}{Topology}{34}{1995}{423-472}
%
\bibitem{LMO}T.~T.~Q.~Le, J.~Murakami, T.~Ohtsuki, {\em A universal
quantum invaraint of 3-manifolds}, preprint, November, 1995.
%
\bimn{LM}{S.~Cappell, R.~Lee, E.~Miller}{A symplectic geometry
approach to generalized Casson's invariants of $3$-manifolds}{Bull.
Amer. Math. Soc.}{22}{1990}{269-275}
%
\bimn{Wi2}{E.~Witten}{Topological quantum field theory}{Commun. Math.
Phys.}{117}{1988}{353-386}
%
\bimn{AJ}{M.~Atiyah, L.~Jeffrey}{Topological Lagrangians and
cohomology}{J. of Geometry and Physics}{7}{1990}{119-136}
%
\bibitem{SeWi2}N.~Seiberg, E.~Witten, {\em Gauge dynamics and
compactification to three dimensions}, preprint IASSNS-HEP-96-78,
RU-96-55, hep-th/9607163.
%
\bibitem{AH}M.~Atiyah, N.~Hitchin, {\em The geometry and dynamics of
magnetic monopoles}, Princeton University Press, 1988.
%
\bibitem{chalmers}G.~Chalmers, A.~Hanany,
{\em Three dimensional gauge
theories and monopoles}, preprint hep-th/9608105.
%
\bibitem{hanany}A.~Hanany, E.~Witten, {\em Type IIB superstrings,
BPS monopoles, and three-dimensional gauge dynamics},
preprint hep-th/9611230.
%
\bibitem{Lescop}C.~Lescop, {\em Global surgery formula for the
Casson-Walker invariant}, Annals of Mathematics Studies, {\bf 140},
Princeton University Press, Princeton, NJ, 1996.
%
\bibitem{topax}M.~Atiyah, {\it The geometry and physics of knots},
Lezioni Lincee, Cambridge University Press, 1990.
%
\bimn{sawin}{S.~Sawin}
{Links, Quantum groups, and  TQFTs}{Bull. Amer.
Math. Soc.}{33}{1996}{413-446}
%
\bimn{At}{M. Atiyah}{On framings of
3-manifolds}{Topology}{29}{1990}{1-7}
%
\bibitem{SSZ}S.~Sethi, M.~Stern, E.~Zaslow, {\em Monopole and dyon
bound states in $N=2$ supersymmetric Yang-Mills theories}, preprint
HUTP-95/A031, DUK-M-95-12, hep-th/9508117.
%
\bimn{BNWi}{D.~Bar-Natan, E.~Witten}{Perturbative expansion of
Chern-Simons theory with non-compact gauge group}{Commun. Math.
Phys.}{141}{1991}{423-440}
%
\bimn{EMSS}{S.~Elitzur, G.~Moore, A.~Schwimmer, N.~Seiberg}{Remarks
on the canonical quantization of the Chern-Simons-Witten
theory}{Nucl. Phys.}{B326}{1989}{108-134}
%
\bibitem{Kol}M.~Kontsevich, {\em Rozansky-Witten invariants via
formal geometry}, preprint dg-ga/9704009.
%
\bibitem{Ka}M.~Kapranov, {\em Rozansky-Witten invariants via Atiyah
classes}, preprint alg-geom/9704009.


\end{thebibliography}
\end{document}